\documentclass{aa}  

\usepackage{graphicx}
\usepackage{txfonts}
\usepackage[colorlinks,allcolors=blue,breaklinks=true]{hyperref}
\usepackage{longtable}
\usepackage{rotating}
\usepackage{lscape}
\usepackage{xcolor}
\usepackage{tcolorbox}
\usepackage{enumitem}
\usepackage{placeins}

\usepackage{bm}

\begin{document}

\title{Data-driven characterisation of variable source light curves with multimodal representation learning}

\author{M. Madarász
        \inst{1,2}
        \thanks{Corresponding author: \texttt{madarasz.mate@csfk.org}}
        \and
        G. Marton
        \inst{1}
        \and
        M. Audard
        \inst{3}
}

\institute{
    HUN-REN Research Centre for Astronomy and Earth Sciences, Konkoly Observatory, MTA Centre of Excellence, Konkoly Thege Mikl\'{o}s \'{u}t 15-17., 1121 Budapest, Hungary
    \and
    Department of Experimental Physics, Institute of Physics, University of Szeged, D\'{o}m t\'{e}r 9, 6720 Szeged, Hungary
    \and
    Department of Astronomy, University of Geneva, Chemin Pegasi 51, 1290 Versoix, Switzerland
}

\date{Received 25 June 2026; accepted 20 October 2026}

\abstract
{Large-scale photometric surveys such as the Zwicky Transient Facility (ZTF) and the Vera C.\ Rubin Observatory Legacy Survey of Space and Time (LSST) are producing light curves for hundreds of millions of variable sources. Characterising these sources has traditionally relied on supervised classifiers trained with hand-crafted features and curated label sets, limiting their applicability to predefined taxonomies and individual surveys.}
{We present a general, survey-agnostic framework for the characterisation of variable sources from photometric time series that operates entirely without labelled training data and produces low-dimensional representations capturing light curve properties such as amplitude, periodicity, and asymmetry directly from the observations.}
{Our method combines two complementary views of each light curve in a multimodal architecture. Raw time-domain observations are processed by a Transformer encoder, and a generalised Lomb--Scargle (GLS) periodogram is processed by a convolutional neural network. Training uses a split-half augmentation strategy, where each light curve is randomly divided into two temporal halves that serve as independent views of the same underlying variability. The two branches are trained jointly using the Variance-Invariance-Covariance Regularisation (VICReg) objective to produce embeddings that are invariant across the two halves while maintaining high information content per dimension. No labels enter the training procedure at any stage.}
{We evaluate the framework on 37\,006 ZTF sources spanning 14 variable star classes across four broad physical categories. A linear classifier trained on the frozen 64-dimensional embeddings achieves 77.1\% balanced accuracy at the individual class level and 90.5\% at the broad category level. Combining the light curve and periodogram branches improves balanced accuracy by up to 14.9 percentage points over the best single modality for individual classes, confirming that the two branches capture complementary information. The embedding encodes interpretable light curve properties on distinct embedding dimensions, and within-class analysis reveals physically interpretable substructure that the reference labels do not resolve. Retraining on ZTF $g$ and $i$ bands without modification reproduces the main results, and concatenating multi-band embeddings improves balanced accuracy to 86.1\%, demonstrating complementary wavelength-dependent information capture.}
{The self-supervised framework produces physically interpretable representations without requiring labelled training data, phase folding, or survey-specific preprocessing. The framework is survey-agnostic by design and directly applicable to other ground-based and space-based photometric surveys with different cadences and wavelength coverage. The trained model and the source code with tutorials are publicly released.}

\keywords{methods: data analysis -- methods: statistical -- stars: variables: general -- techniques: photometric}

   \maketitle
   \nolinenumbers
%
\section{Introduction}\label{sec:introduction}
 
Variable stars are among the most informative objects in time-domain astrophysics, as changes in their brightness directly trace the physical processes that drive their variability, from stellar pulsations and binary interactions to accretion, mass loss, and magnetic activity. The study of stellar variability has been reshaped by wide-field photometric surveys such as the Optical Gravitational Lensing Experiment \citep[OGLE,][]{Udalski2015}, the Zwicky Transient Facility \citep[ZTF,][]{Bellm2019,Masci2019}, and the ESA Gaia mission \citep{GaiaDR3}, which have collectively produced light curves for hundreds of millions of sources. The Vera C.\ Rubin Observatory Legacy Survey of Space and Time \citep[LSST,][]{Ivezic2019} and the upcoming Nancy Grace Roman Space Telescope \citep{Spergel2015} will significantly increase this volume. Gaia Data Release 3 (DR3) alone has delivered epoch photometry for billions of sources \citep{Eyer2023}, with supervised classifiers applied to 12.4 million variable candidates across 25 classes \citep{Rimoldini2023}. These efforts illustrate both the huge scientific potential and the practical challenge of characterising stellar variability in the era of big data, where scalable methods are essential to translate the growing observational volume into physical understanding.
 
The classification of variable stars from light curves has a rich methodological history. Early automated approaches extracted hand-crafted statistical features from the time series, such as period, amplitude, skewness, kurtosis, and various indices describing the shape and regularity of the variability \citep{Debosscher2007,Richards2011,Dubath2011}. Dedicated software libraries such as FATS \citep{Nun2015} standardised the computation of these features and were combined with classical machine learning (ML) classifiers such as random forests \citep[RFs,][]{breiman2001} to classify light curve data. This feature-engineering approach was largely motivated by the variable lengths and irregular temporal sampling of light curve data, which make it difficult to use them directly as inputs to most standard ML algorithms. A practical workaround adopted by several studies was to render each light curve as a two-dimensional image and apply standard convolutional neural networks \citep[CNNs, e.g.][]{Monsalves2024, Szklenar2026}, taking advantage of their proven effectiveness in image classification. While this increases the input dimensionality considerably (from a few hundred observations to tens of thousands of pixels, most of which are background), CNNs can learn to focus on the informative regions and still achieve strong classification results. More naturally suited to sequential data, recurrent neural networks (RNNs), and in particular Long Short-Term Memory \citep[LSTM,][]{Hochreiter1997} networks, were also applied to variable star light curves \citep{Naul2018,JamalBloom2020}, with the advantage of processing observations directly in sequence without a fixed input size. More recently, Transformer architectures \citep{Vaswani2017}, which replace recurrence with multi-head self-attention, have become the state of the art for sequence modelling in astronomy as well, with models such as ASTROMER \citep{DonosoOliva2023} showing strong results on variable star classification. Transformers are well suited to irregularly sampled light curves because their attention mechanism can relate any pair of observations regardless of temporal separation, learning which time scales and observation pairs carry the most information for a given source.
 
These supervised classifiers, whether built on hand-crafted features or deep neural networks, are remarkably effective and can achieve high classification accuracies for well-populated variability classes \citep{Rimoldini2023, SanchezSaez2021, Szklenar2026}, but come with important limitations. If the training labels are noisy or systematically biased, a sufficiently expressive model will reproduce those biases rather than capture the true underlying astrophysics, and may still report high accuracy on test data drawn from the same biased distribution. Performance estimates may therefore not reflect how the classifier behaves on genuinely new data from a different survey or sky region. More fundamentally, supervised classification is limited to the classes defined by the training labels. It can assign a source to one of the predefined categories, but it cannot discover new subgroups or continuous physical structure.
 
Unsupervised and self-supervised learning offer an alternative to supervised classification by learning representations of data without relying on labels, letting the data reveal its own structure. Among the various approaches, autoencoders \citep[AEs,][]{HintonSalakhutdinov2006} and their probabilistic extension, variational autoencoders \citep[VAEs,][]{Kingma2014}, are widely used in variable star studies. They learn a compact latent representation by training a network to compress the input and then reconstruct it. \citet{Naul2018} showed that features from a recurrent autoencoder trained on irregularly sampled light curves can match hand-crafted features for downstream classification, establishing AEs as a viable label-free approach. \citet{ValenzuelaPichara2018} demonstrated that purely unsupervised clustering in a feature space can recover known variability classes without any labels. Most recently, \citet{Huijse2025} trained three VAEs on approximately four million Gaia DR3 sources, combining XP low-resolution spectra, a $\Delta m$/$\Delta t$ scatter-plot representation \citep{Mahabal2017}, and phase-folded light curves into a multimodal latent space that captures astrophysically meaningful structure. While these reconstruction-based approaches have produced valuable results, they face several challenges when applied to diverse and long light curves. The decoder is optimised to reproduce every raw observation including noisy measurements and instrumental artifacts. Phase-folding the light curve before encoding can mitigate this for periodic sources, but requires an accurate period estimate and destroys aperiodic or transient features. Stochastic and transient variables are particularly affected, as the point-wise reconstruction loss averages over their diverse, non-repeating patterns (see Sect.~\ref{sec:disc_comparison} for details). These limitations motivate a reconstruction-free alternative, which we describe in the following.
 
Contrastive representation learning \citep{Chen2020} offers an alternative that does not require point by point reconstruction, training the network instead to produce similar representations for different augmented views of the same source. This paradigm has been applied to light curves e.g. by \citet{Ding2026}, who use partially overlapping temporal subsequences as views, and in a multimodal image-based setting by \citet{RizhkoBloom2025}, who align light curves, spectra, and metadata through a Contrastive Language Image Pre-training (CLIP)-based objective. Our approach differs in two respects. Rather than selecting contiguous temporal crops that may miss important events or reduce phase coverage, we randomly assign each light curve's observations into two subsets that independently span the full observing baseline (see Fig.~\ref{fig:pipeline}). This ensures that both views retain the full statistical properties of variability, including the characteristic frequency and amplitude of events such as bursts, dips, and eclipses. We also provide frequency-domain information through a separate periodogram branch, allowing the model to capture both temporal and frequency-domain structure. Because the network has limited capacity, it cannot memorise individual sources and is pushed toward learning shared variability features, organising the embedding space by variability type. To our knowledge, this split-half temporal augmentation has not been previously applied to light curve representation learning. The strategy turns the properties that challenge reconstruction-based methods into advantages, as longer light curves provide more observations per half, noisy points act as natural data augmentation, and no phase folding is required.
 
We evaluate the framework on a catalogue of 37\,006 ZTF sources spanning 14 variable classes assembled from the Automatic Learning for the Rapid Classification of Events broker \citep[ALeRCE,][]{Forster2021,SanchezSaez2021}. The paper is organised as follows. Section~\ref{sec:data} describes the construction of the multi-class catalogue and the light curve preprocessing pipeline. Section~\ref{sec:method} presents the architecture, the split-half augmentation strategy, the training objective, and the evaluation framework. Section~\ref{sec:results} reports classification performance, the contributions of individual input modalities, the light curve properties encoded in the embedding, and multi-band generalisation. Section~\ref{sec:discussion} discusses class confusions and label quality, within-class structure, compares our approach with reconstruction-based methods, and discusses multi-band complementarity. Section~\ref{sec:conclusions} summarises our findings.

\section{Data}\label{sec:data}

\subsection{Source catalogue construction}\label{sec:catalogue}
 
To evaluate our framework across a broad range of stellar variability, we constructed a multi-class catalogue from the ALeRCE broker \citep{Forster2021}, which processes the full ZTF alert stream and classifies millions of variable and transient sources using a hierarchical balanced RF trained on ZTF light curve features \citep{SanchezSaez2021}. We chose ALeRCE as our main source of class labels for several reasons. First, all classifications are derived from ZTF photometric data, ensuring that the labels correspond to the same survey and observing conditions as the light curves planned to be used in our study. Second, the classifier was applied uniformly to all sources using the same methodology, feature set, and selection criteria. While dedicated catalogues from individual studies may provide more reliable labels for specific classes, combining labels from many sources would introduce inconsistent selection functions and labelling criteria, and deciding which catalogue to trust for each class is itself a non-trivial task. Third, ALeRCE provides probabilistic classifications across a wide taxonomy, allowing us to select high-confidence sources and control the class composition. This approach is analogous to that of \citet{Huijse2025}, who used the Gaia DR3 supervised classification \citep{Rimoldini2023} as reference labels for evaluating their unsupervised study.
 
We queried sources classified by ALeRCE with a best-class probability of at least $p = 0.5$, using the light curve classifier \texttt{lc\_classifier\_BHRF\_forced\_phot}\footnote{ALeRCE classifications queried on 28 May 2026. As ALeRCE is an active broker, source counts may differ slightly at later dates.}. Table~\ref{tab:catalogue} lists the full set of classes considered. We retained the following classes for our study: EA (detached eclipsing binaries), EB/EW (contact and semi-detached eclipsing binaries), RRab (RR~Lyrae fundamental mode), RRc (RR~Lyrae first overtone), CEP (Cepheids), DSCT ($\delta$~Scuti variables), LPV (long-period variables), RS~CVn (RS Canum Venaticorum variables), QSO (quasi-stellar objects), AGN (active galactic nuclei), Blazar, YSO (young stellar objects), and CV/Nova (cataclysmic variables and novae). Classes with fewer than 200 sources at $p > 0.5$ were excluded from the study, as their small sample sizes would not support reliable evaluation. This criterion removed Periodic-Other, Microlensing, SESN, SNII, SNIIn, SLSN, and TDE. For classes where the query returned more than 5000 sources (EB/EW, RRab, RRc, LPV, QSO, and CV/Nova), we retained the 5000 highest-probability sources to limit class imbalance.

\begin{table}[!ht]
  \caption{Source catalogue summary.}
  \label{tab:catalogue}
  \centering
  \begin{tabular}{l r r r r}
    \hline\hline
    Class & $N_{p>0.5}$ & $N_\mathrm{used}$ & Fraction & Balanced \\
    \hline
    QSO     & 17\,334 & 4\,729 & 12.8\% & 10.2\% \\
    RRab   & 34\,963 & 4\,606 & 12.4\% & 10.0\% \\
    RRc    & 13\,355 & 4\,247 & 11.5\% & 9.6\% \\
    EB/EW   & 35\,109 & 4\,140 & 11.2\% & 9.5\% \\
    CV/Nova & 6\,663  & 4\,098 & 11.1\% & 9.5\% \\
    LPV     & 31\,770 & 3\,966 & 10.7\% & 9.3\% \\
    EA      & 3\,829  & 2\,904 & 7.8\%  & 8.0\% \\
    AGN     & 2\,196  & 1\,935 & 5.2\%  & 6.5\% \\
    DSCT    & 2\,575  & 1\,863 & 5.0\%  & 6.4\% \\
    SN~Ia    & 3\,592$^\dagger$ & 1\,662 & 4.5\% & 6.0\% \\
    YSO     & 1\,268  & 1\,214 & 3.3\%  & 5.1\% \\
    CEP     & 1\,522  & 1\,005 & 2.7\%  & 4.7\% \\
    RS~CVn   & 481     & 355    & 1.0\%  & 2.8\% \\
    Blazar  & 282     & 282    & 0.8\%  & 2.5\% \\
    \hline
    Total   &         & 37\,006 & 100\% & 100\% \\
    \hline
  \end{tabular}
  \tablefoot{$N_{p>0.5}$: number of sources available from ALeRCE at best-class probability $p > 0.5$. $N_\mathrm{used}$: number of unique sources retained after the 5000-source cap, light curve retrieval, deduplication, and preprocessing (Sect.~\ref{sec:preprocessing}) with a minimum of 50 observations. Fraction: natural class distribution based on $N_\mathrm{used}$. Balanced: effective class distribution under sqrt-inverse sampling during training (Sect.~\ref{sec:implementation}). $^\dagger$SN~Ia sources are from the ZTF SN~Ia~DR2 \citep{Rigault2025}, not ALeRCE.}
\end{table}
 
Type~Ia supernovae represent a fundamentally different kind of variability compared to the periodic and stochastic classes listed above. Their light curves follow a characteristic template shape, rising rapidly to a peak brightness and then fading over weeks, making them transient events rather than persistent variables. The ALeRCE catalogue contained only 508 SN~Ia sources at $p > 0.5$, which we considered insufficient for robust representation. To test whether our framework can capture this distinct light curve morphology alongside the other variability types, we therefore drew SN~Ia sources from the second data release of the ZTF Cosmology Science Working Group \citep[ZTF SN~Ia~DR2,][]{Rigault2025}, which provides 3592 spectroscopically confirmed SN~Ia with ZTF photometry. From this catalogue, we used only the sky coordinates and the SN~Ia class label.
 
For downstream evaluation, we defined three levels of class hierarchy. The finest level (\texttt{class}) uses all 14 individual classes listed in Table~\ref{tab:catalogue}. The intermediate level (\texttt{class\_mid}) merges AGN, Blazar, and QSO into a single AGN class, as these represent the same physical phenomenon observed under different conditions, resulting in 12 classes. The broadest level (\texttt{class\_broad}) groups the 14 classes into four physical categories: pulsating (RRab, RRc, CEP, DSCT, LPV, RS~CVn), geometric (EA, EB/EW), stochastic (QSO, AGN, Blazar, YSO, CV/Nova), and transient (SN~Ia). We include RS~CVn in the pulsating category because their smooth, quasi-sinusoidal light curves from rotational spot modulation are more similar to pulsating variables than to the sharp eclipses of geometric variables.
 
\subsection{Light curve data}\label{sec:lcdata}

\subsubsection{Light curve retrieval}\label{secc:lcretrieval}

Rather than using the pre-processed light curves available from ALeRCE, we downloaded raw ZTF light curves from the NASA/IPAC Infrared Science Archive \citep[IRSA,][]{Masci2019} using the ZTF Light Curve 
API\footnote{\url{https://irsa.ipac.caltech.edu/docs/program_interface/ztf_lightcurve_api.html}}. For each source, we queried the ZTF Data Release 24 photometry data using the ALeRCE-provided RA and Dec coordinates. ZTF provides photometry in three optical bands ($g$, $r$, and $i$). We adopt the $r$ band for the main analysis because it yields the largest dataset (37\,006 sources after preprocessing, compared to 35\,737 in $g$ and 17\,978 in $i$) and the highest median number of observations per source (679, compared to 464 in $g$ and 152 in $i$). However, we also evaluate the final framework on the other two bands to test wavelength independence. We chose to work with uncleaned archival data rather than preprocessed light curves for two reasons. First, we wanted to evaluate the robustness of our framework on data that contains the full range of observational artifacts present in survey photometry, including outliers, bad measurements, and instrumental artifacts. Many traditional approaches and some ML methods are sensitive to such artifacts, but our split-half augmentation strategy naturally mitigates their effect by treating noisy observations as a form of data augmentation (see Sect.~\ref{sec:method}). Second, using the full archival data maximises the number of observations per source, which directly benefits the split-half approach by providing more diverse temporal subsets for training. Not all queried sources yielded usable light curves, due to missing ZTF $r$-band counterparts and deduplication. Together with the minimum observation threshold of Sect.~\ref{sec:preprocessing}, these steps account for the reduction from $N_{p>0.5}$ to $N_\mathrm{used}$ in Table~\ref{tab:catalogue}.
 
\subsubsection{Preprocessing}\label{sec:preprocessing}

Our publicly released code implements several preprocessing strategies for photometric light curves, including flag-based filtering, global and local sigma clipping with configurable thresholds, and point aggregation with multiple modes. Users can enable any combination of these depending on their data and science case. For the ZTF data used in this study, we applied only what we term magnitude-grid nightly aggregation with a grid width of 0.01~mag, and deliberately did not apply anything else.

Magnitude-grid nightly aggregation groups same-night observations by magnitude into bins of a fixed width, retaining one observation per occupied bin. For constant or slowly varying sources, intra-night observations fall within one or two bins and are effectively averaged, producing a more uniformly sampled light curve across the full observing baseline. For rapidly varying sources such as short-period pulsators with significant intra-night amplitude, multiple bins are occupied and several observations per night are preserved, retaining the astrophysical signal. Figure~\ref{fig:maggrid} in Appendix~\ref{app:preprocessing} illustrates two cases where aggregation corrects an aliased period for an RR~Lyrae variable and removes 12\% of redundant intra-night scatter for a $\delta$~Scuti variable without loss of phase-coverage or signal. The latter source has a peak-to-peak amplitude of only 0.2~mag, confirming that the 0.01~mag grid width preserves the waveform shape even for low-amplitude, short-period variables.

We selected the grid width of 0.01~mag based on a period recovery analysis on 25\,088 pulsating variables from \citet{Szklenar2026} with known OGLE reference periods, where aggregation improved exact period recovery from 73.7\% to 77.5\% compared to the raw ZTF data. The full comparison across grid widths is presented in Appendix~\ref{app:preprocessing}.

After preprocessing, sources with fewer than 50 observations were excluded, following the minimum threshold commonly adopted for reliable variability and periodicity analysis of ZTF light curves \citep{Coughlin2021, Healy2024}. This final filtering step produces the source counts listed as $N_\mathrm{used}$ in Table~\ref{tab:catalogue}, totalling 37\,006 sources with a median of 679 $r$-band observations per source (range: 50--1966).

\subsection{Dataset splits}\label{sec:splits}

We divided the dataset into three non-overlapping subsets using stratified random sampling on the 14-class labels, ensuring proportional class representation in each subset.

The training set (70\%, 25\,906 sources) is used to adjust model parameters through the optimisation objective of Sect.~\ref{sec:vicreg}. Because the model parameters were adjusted based on these sources, they cannot be used to reliably assess generalisation performance. The validation set (15\%, 5\,550 sources) is evaluated after each training epoch to monitor convergence and detect overfitting. Overfitting occurs when the model memorises source-specific features rather than learning general variability patterns, appearing as a divergence between decreasing training loss and increasing validation loss. Training is stopped where the validation loss is lowest, before divergence occurs. During the hyperparameter search, validation metrics also guide the selection of loss weights, architectural choices, and training parameters. The test set (15\%, 5\,550 sources) is a separate, independent subset held out entirely from both training and model selection, with no overlap with the training or validation sets. Because no decisions were made based on these sources, the test set provides an unbiased estimate of how the trained model performs on completely unseen data, reflecting what a user of the released model can expect when applying it to new light curves. We emphasise that class labels are used only for constructing the stratified split and for final evaluation. They never enter the training objective at any stage.

\section{Method}\label{sec:method}
 
Using the light curves downloaded and preprocessed as described in Sects.~\ref{secc:lcretrieval} and~\ref{sec:preprocessing}, we further prepare each source for training through a series of augmentation and input preparation steps (Sect.~\ref{sec:pipeline}). The processed inputs are fed into a two-branch neural network (Sect.~\ref{sec:architecture}) that independently encodes the time-domain and frequency-domain views of each source, then fuses them into a compact, fixed-size embedding vector. This embedding is optimised with a contrastive learning objective (Sect.~\ref{sec:vicreg}) that encourages the model to produce similar representations for different temporal subsets of the same source, without access to any class labels during training. The result is a learned representation where proximity reflects physical similarity in variability properties, and where distinct variability morphologies can emerge as coherent groups purely from the structure of the data. We evaluate whether this is the case by training simple classifiers on the frozen embeddings and measuring how well they recover known variability classes (Sect.~\ref{sec:evaluation}).
 
\subsection{Processing pipeline}\label{sec:pipeline}

\begin{figure*}
  \centering
  \includegraphics[width=\textwidth]{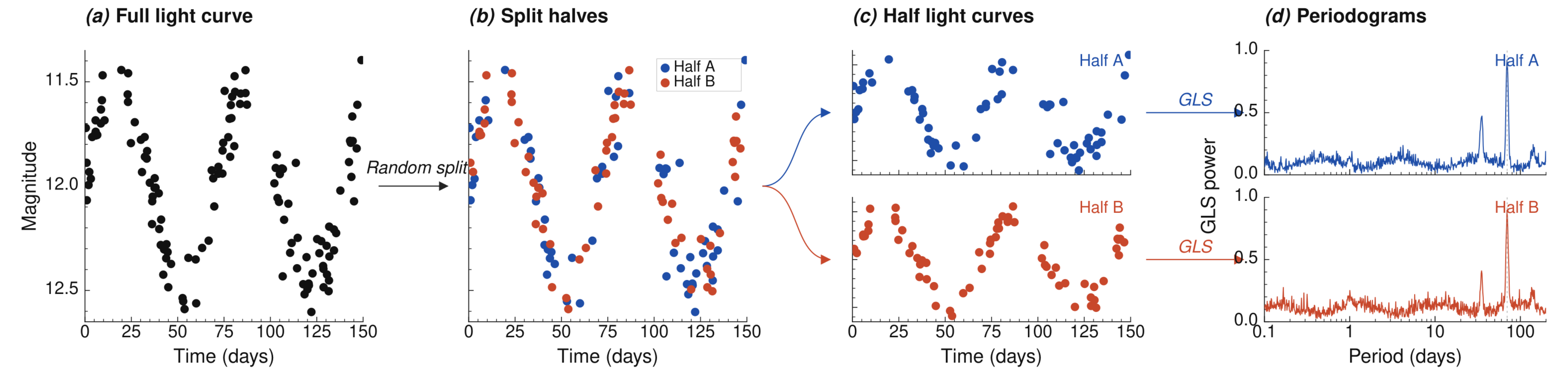}
 \caption{Split-half augmentation and periodogram computation pipeline. \textit{(a)} A synthetic light curve of a periodic variable. \textit{(b)} Observations randomly assigned to two halves (blue: Half~A, red: Half~B). \textit{(c)} The two halves as independent partial light curves. \textit{(d)} Independent GLS periodograms computed on each half.}
  \label{fig:pipeline}
\end{figure*}

The processing pipeline transforms each preprocessed light curve into four network-ready arrays: two partial time-series and two periodograms. Figure~\ref{fig:pipeline} illustrates the full sequence for a periodic variable. The light curve is first standardised in time and magnitude (Sect.~\ref{sec:normalisation}). The observations (panel~a) are then randomly assigned to two halves of roughly equal size (panel~b), producing two independent partial light curves that span the same observing baseline (panel~c, Sect.~\ref{sec:splithalf}). Each half is independently downsampled to a random length to prevent the model from using observation count as a shortcut feature (Sect.~\ref{sec:lengthaug}). Finally, a GLS periodogram is computed independently on each half from the raw magnitudes, producing a fixed-length frequency-domain representation alongside each partial time-series (panel~d, Sect.~\ref{sec:periodogram_computation}). The result is two complementary views of the same source, each consisting of a time-domain light curve and a frequency-domain periodogram, which are processed by separate branches of the neural network described in Sect.~\ref{sec:architecture}. The following subsections describe each step in detail.
 
\subsubsection{Light curve standardisation}\label{sec:normalisation}
 
Neural networks are sensitive to the scale and range of their inputs, and training is most stable when all input features occupy a comparable numerical range centred near zero. Before splitting, we therefore standardise each light curve in both time and magnitude. Timestamps are linearly mapped to the $[0,1]$ interval so that the model learns from the relative temporal spacing of observations rather than from absolute calendar dates. For magnitude standardisation, the publicly released code supports two modes: per-sample and global standardisation. In both cases, the per-source median magnitude is subtracted, centering all light curves around zero and thus removing the apparent brightness level. In per-sample mode, the median-subtracted residuals are further divided by the median absolute deviation (MAD) of each source individually. This scales every light curve to unit spread, forcing the model to focus on the shape of the variability rather than its amplitude. In global mode, the residuals are instead divided by a shared scale factor computed as the 99th percentile of the absolute deviations across all training sources. This preserves relative amplitude differences between sources, for instance, a 2-mag eclipsing binary retains a signal roughly 20 times larger than a 0.1-mag rotational variable. Because variability amplitude is itself a physically meaningful and discriminative property, we adopted global standardisation.
 
The model receives two features per observation, standardised time and magnitude. We excluded photometric uncertainties because preliminary experiments showed the network exploits them as a shortcut, producing embeddings that correlate with apparent brightness rather than variability properties.
 
\subsubsection{Split-half augmentation}\label{sec:splithalf}
 
The core augmentation strategy is a random temporal splitting of each light curve (Fig.~\ref{fig:pipeline}). At every training epoch, the $n$ observations of a given source are randomly permuted and divided into two subsets of $\lfloor n/2 \rfloor$ and $n - \lfloor n/2 \rfloor$ observations, respectively. Each subset is then sorted back into temporal order, producing two partial light curves that span the same observing baseline but sample it at different times. Because the split is random rather than sequential, both halves contain observations from all epochs of the survey, ensuring each half independently samples the full range of variability exhibited by the source.

For periodic variables, both halves sample the same underlying phase curve, so a model that captures the waveform shape and period will naturally produce similar representations for the two halves. For stochastic variables such as certain YSOs or AGNs, the two halves share the same long-term statistical properties (characteristic timescales, amplitude distributions, statistical aspects of bursts or dips) even though the specific events might differ between halves. The training objective (Sect.~\ref{sec:vicreg}) encourages the model to match the representations of the two halves, forcing it to learn general variability properties that are common to both rather than specific observations in either half.
 

Recent self-supervised approaches to light curve representation learning have employed random masking \citep[e.g.][]{DonosoOliva2023} or overlapping temporal crops \citep[e.g.][]{Ding2026} as augmentation strategies. Masking produces views that share most observations, limiting augmentation diversity, while cropping selects contiguous segments that may exclude important events or sample only a fraction of periodic phases. The split-half augmentation instead ensures that both views independently span the full observing baseline, sampling the complete range of timescales and variability behaviour. To our knowledge, this approach has not been previously applied to light curve representation learning.
 
\subsubsection{Length randomisation}\label{sec:lengthaug}
 
Without further intervention, the two halves always contain exactly $\lfloor n/2 \rfloor$ and $n - \lfloor n/2 \rfloor$ observations, making sequence length a deterministic function of the total observation count. The model could exploit this as a shortcut rather than learning variability features. To break this, each half is independently downsampled to a random fraction between 20\% and 100\% of its length, so the two halves end up with different numbers of observations. This forces the model to rely on variability properties rather than observation count for matching.
 
\subsubsection{Periodogram computation}\label{sec:periodogram_computation}
 
For each half of the split light curve, we compute an independent GLS \citep{ZechmeisterKurster2009} periodogram using the NUFFT-accelerated implementation \texttt{nifty-ls}\footnote{\url{https://github.com/flatironinstitute/nifty-ls/}} \citep{Garrison2024}, which provides order-of-magnitude speedups over traditional methods. The periodograms are computed from the pre-standardisation magnitudes of each half, because the GLS formalism benefits from data in physical units and can use photometric uncertainties for inverse-variance weighting when available. The resulting periodogram captures the frequency-domain structure of the variability as seen by each half independently.
 
We search a period range from 0.03 to 1000~days, spanning the full range of variability timescales in our dataset from $\delta$~Scuti pulsations at the short end to long-period variables at the long end. The GLS is first evaluated on a dense linear-frequency grid. This oversampling is necessary because peaks in the periodogram have a natural width of approximately $1 / T_\mathrm{baseline}$ in frequency space, and a coarse grid would miss narrow peaks from short-period variables such as $\delta$~Scuti stars and RR~Lyrae variables. The dense periodogram is then compressed into 1000 logarithmically spaced period bins by retaining the maximum power within each bin, reducing it to a size that the CNN can process efficiently while preserving the dominant period peak and its harmonics. Each periodogram is then standardised by subtracting and dividing by the median power, placing the noise floor near zero so that periodograms from sources with different baselines and sampling patterns are directly comparable.

During development we fixed this grid at 300\,000 points, an oversampling generous enough that frequency resolution never limited any architectural or training choice. After the model was fixed by the label-free selection of Sect.~\ref{sec:evaluation}, a targeted experiment on the training and validation set showed that 200\,000 points reproduce the 300\,000-point classification accuracy, and we adopt this value for the released model. A coarser 100\,000-point grid costs less than one point of accuracy at roughly half the training time, a useful lighter-weight alternative. The full sweep and timings are given in Appendix~\ref{app:ndense}.

 
This approach differs fundamentally from the common practice of extracting only the peak period from the periodogram and using it to phase-fold the light curve \citep{Huijse2025, Szklenar2026}. Phase folding discards all periodogram information beyond the dominant period, including harmonics, secondary periods, and the overall power spectrum shape. Our approach retains the full periodogram as a 1000-bin CNN input, preserving this information, while leaving the light curve in its original temporal form so that the time-domain branch can capture aperiodic and stochastic variability that folding would destroy. The two modalities are complementary. For periodic sources, the periodogram captures the dominant period and its harmonics while the light curve captures the waveform shape. For stochastic sources, a flat, noise-dominated periodogram is itself a discriminative signal that helps separate periodic from aperiodic classes. The two halves of each source produce slightly different periodograms with the same dominant peaks but different noise realisations, providing natural augmentation for the frequency-domain branch alongside the time-domain augmentation from the split itself.

\subsection{Architecture}\label{sec:architecture}

\begin{figure}[!ht]
  \centering
  \includegraphics[width=0.8\columnwidth]{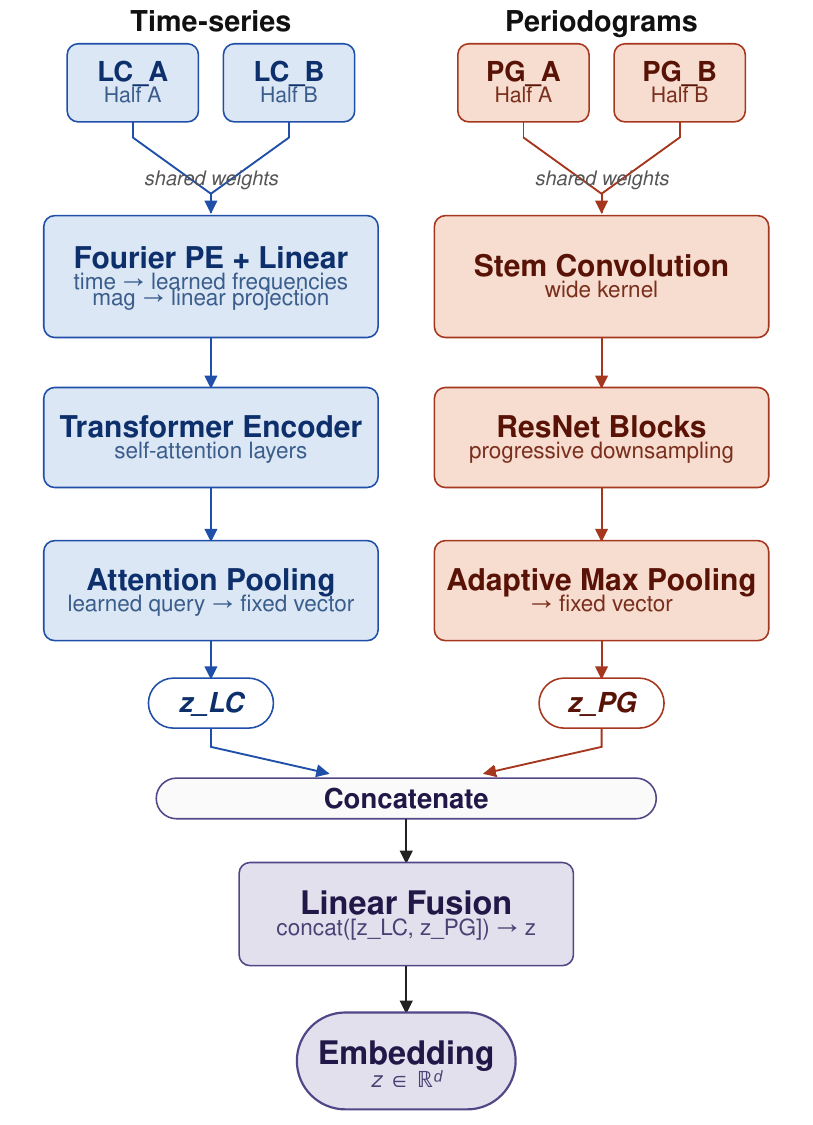}
\caption{Two-branch encoder architecture. The four inputs from the processing pipeline are processed by shared-weight encoders: a Transformer for the time-series (left, Sect.~\ref{sec:lc_encoder}) and a ResNet for the periodograms (right, Sect.~\ref{sec:pg_encoder}). The branch outputs are concatenated and fused into the final embedding $\mathbf{z} \in \mathbb{R}^d$ (Sect.~\ref{sec:fusion}).}
  \label{fig:architecture}
\end{figure}

The network consists of two encoder branches that process the time-domain and frequency-domain inputs independently, followed by a fusion layer that combines them into a single embedding vector. Figure~\ref{fig:architecture} illustrates the full architecture. The four inputs from the processing pipeline (two partial light curves and two periodograms from the same source) enter at the top and flow downward through two parallel branches. The left branch uses a Transformer encoder \citep{Vaswani2017} to process the variable-length, irregularly sampled time-series, while the right branch uses a residual convolutional network \citep[ResNet,][]{He2016} to process the fixed-length power spectrum. Both branches produce fixed-size output vectors, which are concatenated and projected into the final embedding through a single fusion layer, a linear map followed by a GELU activation, at the bottom of the diagram. The two halves of each source are processed by the same shared-weight encoders, producing two embeddings per source that are compared by the training objective (Sect.~\ref{sec:vicreg}). Each component is described below.

\subsubsection{Light curve encoder}\label{sec:lc_encoder}

The light curve encoder, shown on the left side of Fig.~\ref{fig:architecture}, must handle two challenges, irregular temporal sampling and variable sequence lengths. Transformer networks \citep{Vaswani2017} are naturally suited to both, because their core operation, self-attention, computes a weighted similarity score between every pair of observations in the sequence, allowing each data point to gather information from all others regardless of their temporal separation or position in the input.

Each observation enters the encoder as a pair of standardised values, time and magnitude. To represent temporal information, we use learnable Fourier positional encoding \citep{Li2021}. The standardised timestamp is passed through a set of learned frequency and phase parameters, producing a representation via sine and cosine functions. The frequencies and phases are learned during training rather than fixed as in the original Transformer. The magnitude is separately projected through a linear layer to the same dimensionality, and the two representations are summed to form the input token for each observation.

The sequence of input tokens is processed by a stack of Transformer encoder layers, each applying multi-head self-attention followed by a feed-forward network with GELU activations \citep{Hendrycks2016}. Self-attention allows every observation to directly attend to every other observation in the light curve, enabling the network to compare brightness levels across the full temporal baseline to assess periodicity, or relate the depth and duration of events at different times. Multiple attention heads capture different types of relationships simultaneously, from short-term correlations to long-term patterns.

After the Transformer layers, we compress the variable-length sequence into a single fixed-size vector using learned attention pooling \citep{Lee2019}, as shown at the bottom of the left branch in Fig.~\ref{fig:architecture}. A single learned query vector attends to all observation representations through multi-head attention, allowing the network to learn which observations are most informative and weight them accordingly. This produces a fixed-size output vector regardless of the input sequence length.

\subsubsection{Periodogram encoder}\label{sec:pg_encoder}

The right side of Fig.~\ref{fig:architecture} shows the periodogram encoder. Unlike the irregularly sampled light curve, the periodogram is a fixed-length one-dimensional signal on a regular grid, making it well suited to a standard CNN architecture. We use a one-dimensional ResNet \citep{He2016}.

The network begins with a stem convolution using a wide kernel, giving the first layer a broad receptive field that can capture local structures such as a dominant period peak and its nearby harmonics. The stem is followed by a series of residual blocks, each halving the spatial dimension through strided convolution. As the signal passes through successive layers, the effective receptive field grows, allowing deeper layers to capture relationships between increasingly distant features, such as the spacing between a fundamental period and its harmonics.

We use Group Normalisation \citep[GroupNorm,][]{Wu2018} rather than the more common Batch Normalisation (BN). For periodic variables, the periodograms within a batch are dominated by a flat noise floor with only a few prominent peaks, making them structurally similar. BN computes statistics across the batch, which would average over these flat profiles and wash out the peak differences that distinguish sources. GroupNorm computes statistics within each sample independently, preserving these differences. At the output, we apply adaptive max pooling to compress the feature maps into a single value per channel. Max pooling is preferred over mean pooling because the discriminative information in a periodogram is concentrated in the peaks. Mean pooling would dilute the peak signal across hundreds of uninformative noise bins, whereas max pooling ensures that detected peaks drive the final representation.

\subsubsection{Fusion}\label{sec:fusion}

At the bottom of Fig.~\ref{fig:architecture}, the fixed-size output vectors from both branches are concatenated and passed through a single linear layer followed by a GELU activation to produce the final embedding. This fusion step allows the network to combine information from both modalities when arranging sources in the embedding space. For example, two sources with similar periodograms but different light curve shapes (such as an RR~Lyrae and a contact binary at similar period) can be separated based on the time-domain features, while sources with similar morphology but different frequency content are separated by the periodogram branch. For aperiodic and stochastic sources, the periodogram branch contributes a noise-dominated power spectrum, which is itself a learnable feature the fusion layer can use to distinguish them from periodic classes.

We found in preliminary architecture experiments described in Appendix~\ref{app:architecture_experiments} that using the encoder outputs directly and fusing them with a single layer consistently outperformed more complex alternatives with additional non-linear layers, which tended to enable memorisation without improving the quality of the learned representations. The published code supports flexible projection and fusion configurations, but we recommend this minimal design.

\subsection{Training objective}\label{sec:vicreg}

The network is trained using a self-supervised contrastive objective that requires no class labels. The core idea is straightforward: the two halves of the same source should produce similar embeddings, because they represent different temporal samples of the same underlying variability. However, simply minimising the distance between paired embeddings leads to a trivial solution where the network maps all sources to a single point. The challenge is to encourage matching within each source while ensuring that the embedding space remains informative and well-structured across the full dataset. We address this using VICReg \citep[Variance-Invariance-Covariance Regularisation,][]{Bardes2022}, a loss function composed of three complementary terms that jointly prevent collapse and encourage a rich, non-redundant representation. 

\subsubsection{VICReg loss}\label{sec:vicreg_loss}

The invariance term pulls together the embeddings of the two halves of each source. Given a batch of $B$ sources with embedding pairs $(\mathbf{z}_a^i, \mathbf{z}_b^i)$ of dimensionality $d$, the invariance loss is the mean squared error between matched pairs:
\begin{equation}\label{eq:invariance}
  \mathcal{L}_\mathrm{inv} = \frac{1}{Bd} \sum_{i=1}^{B} \| \mathbf{z}_a^i - \mathbf{z}_b^i \|^2 .
\end{equation}
Minimising this term encourages the network to produce representations that are stable across different temporal subsets of the same source, pushing sources with similar variability behaviour toward shared regions of the embedding space.

The variance term prevents the embedding from collapsing to a trivial constant. It requires the standard deviation of each embedding dimension across the batch to remain above a target threshold $\gamma$:
\begin{equation}\label{eq:variance}
  \mathcal{L}_\mathrm{var} = \frac{1}{d} \sum_{j=1}^{d} \max\!\bigl(0,\; \gamma - \sigma(\mathbf{z}_j)\bigr) ,
\end{equation}
where $\sigma(\mathbf{z}_j) = \sqrt{\mathrm{Var}(\mathbf{z}_j) + \epsilon}$ is the regularised standard deviation of the $j$-th dimension computed across the batch, and the term is applied independently to both halves and summed. Without this constraint, the invariance term alone would drive all embeddings to a single point, eliminating any useful structure.

The covariance term decorrelates the embedding dimensions pairwise. It penalises the off-diagonal elements of the covariance matrix $\mathbf{C}$ of the batch embeddings:
\begin{equation}\label{eq:covariance}
  \mathcal{L}_\mathrm{cov} = \frac{1}{d} \sum_{j \neq k} C_{jk}^2 ,
\end{equation}
where $C_{jk}$ is the $(j,k)$ entry of the covariance matrix computed from the centred embeddings, applied independently to both halves and summed. Without this term, multiple dimensions could encode the same information (such as amplitude or period). Pushing off-diagonal covariances toward zero encourages each dimension to capture a different aspect of the variability.

The total loss is a weighted sum of the three terms:
\begin{equation}\label{eq:vicreg_total}
  \mathcal{L} = \lambda_\mathrm{inv} \, \mathcal{L}_\mathrm{inv} + \lambda_\mathrm{var} \, \mathcal{L}_\mathrm{var} + \lambda_\mathrm{cov} \, \mathcal{L}_\mathrm{cov} .
\end{equation}
We use $\lambda_\mathrm{inv} = 5$, $\lambda_\mathrm{var} = 25$, $\lambda_\mathrm{cov} = 5$, and $\gamma = 0.2$. The loss is computed in a combined mode, both on the fused embedding and independently on each branch embedding, with the three resulting losses summed with equal weight, so that each branch develops a well-structured representation within its own modality. The rationale for these choices is discussed in Appendix~\ref{app:vicreg_hyperparameters}.

\subsection{Evaluation framework}\label{sec:evaluation}

Class labels (Table~\ref{tab:catalogue}) are used only for evaluation and never enter the training objective at any stage.

\subsubsection{Classification on frozen embeddings}\label{sec:eval_classification}

To evaluate what the learned embeddings encode, we train simple classifiers on the frozen representations produced by the trained model. At inference time, each source is passed through the network once using its full light curve and a periodogram computed from it, without the training-time split, producing a single embedding vector per source. If a simple linear classifier can recover the main variability classes from these fixed representations, it demonstrates that the self-supervised training has organised the embedding space so that different variability types occupy distinct, well-separated regions, without the model ever being told which classes exist.

We train two classifiers with balanced class weights: a logistic regression (LR) and a support vector machine with a radial-basis-function kernel \citep[SVM,][]{Cortes1995}. The logistic regression directly tests linear separability. High accuracy means that different variability types are arranged into regions that can be separated by hyperplanes, so a linear decision boundary suffices. The SVM can additionally capture non-linear structure in the embedding space. When both classifiers achieve similar accuracy, the representation has already done most of the organisational work, and no complex non-linear classifier is needed to extract the variability information. We evaluate classification at the three hierarchical levels defined in Sect.~\ref{sec:catalogue}. Balanced accuracy, the mean per-class recall, is the primary metric, with macro~F1 as a secondary measure.

\subsubsection{Label-free development and label-guided selection}\label{sec:eval_labels}

The model was developed entirely without class labels. All architectural choices, including the encoder branch designs, the attention configuration, the number of layers, the fusion strategy, and the projection dimensions, as well as the VICReg loss weights and training hyperparameters, were made using only the VICReg validation loss and visual inspection of Uniform Manifold Approximation and Projection (UMAP) \citep{McInnes2018} projections. These label-free criteria confirm that the embedding is well-conditioned, maintaining sufficient per-dimension variance and low inter-dimension redundancy, but they cannot reveal whether the learned structure aligns with astrophysical variability classes. The rationale behind these architectural and training decisions is detailed in Appendix~\ref{app:model_development}.

Once the model design was fixed, a small number of remaining choices required evaluating what the embeddings actually encode. The dense periodogram grid resolution described in Sect.~\ref{sec:periodogram_computation}, the embedding dimensionality reported in Appendix~\ref{app:dimension}, and whether both input modalities are necessary as presented in Sect.~\ref{sec:res_modality}. For these comparisons, we introduced class labels through five-fold stratified cross-validation \citep[5-fold CV,][]{Kohavi1995} on the combined training and validation set. Labels do not enter the training objective in any of these experiments. They serve only to compare otherwise identical model variants against each other, selecting the configuration whose embedding structure best captures the variability classes. The embedding itself is still constructed without labels. This means we use the labels to find which configuration produces the most astrophysically meaningful structure, not to optimise the structure toward them.

The test set described in Sect.~\ref{sec:splits} is reserved for a single final evaluation of the selected model. Because no design or selection decision was based on the test sources, it provides an unbiased estimate of how the method generalises to unseen data. Comparable performance between the cross-validation and test-set evaluations confirms that the label-guided selection did not introduce overfitting, and that users of the released model can expect similar performance on new light curves.

\subsubsection{Light curve property analysis}\label{sec:eval_properties}

Beyond discrete class recovery, we assess which continuous light curve properties the embedding captures by correlating each embedding dimension with source-level quantities using the Spearman rank coefficient. Variability amplitude measures the overall strength of the brightness variations. The quartile ratio $Q_{31}$ \citep{Cabral2018} provides a robust measure of the magnitude spread. The GLS period is the dominant period from the periodogram computed on the preprocessed light curve. The autocorrelation length \citep{Cabral2018} measures the characteristic timescale over which consecutive observations remain correlated. The periodicity index $Q$ \citep{Cody2014} quantifies how well a source's variability is described by a periodic model when phase-folded on the best-fit period, ranging from zero for strictly periodic sources to one for irregular or aperiodic behaviour. The asymmetry index $M$ \citep{Cody2014} measures the asymmetry of the magnitude distribution, distinguishing sources dominated by brightening events from those dominated by fading events. The skewness of the magnitude distribution \citep{Cabral2018} captures a complementary aspect of light curve shape. Finally, the number of observations per source ($n_\mathrm{final}$) is included as a diagnostic, because a strong correlation would suggest that the model exploits observation count as a shortcut rather than learning variability properties, which the length augmentation described in Sect.~\ref{sec:lengthaug} is designed to reduce.

We complement the correlation analysis with two-dimensional UMAP projections of the embedding space, coloured by class labels and by the light curve properties above, to visually assess whether the embedding structure aligns with known astrophysical groupings and gradients.

\subsection{Implementation details}\label{sec:implementation}

To improve class imbalance during training, we use a weighted random sampler with sqrt-inverse class weights. Each source in class $c$ receives sampling weight $1/\sqrt{N_c}$, where $N_c$ is the number of training sources in that class. This moderately upweights rare classes (Table~\ref{tab:catalogue}, Balanced column) without the extreme oversampling that full inverse weighting would produce. The final model produces a 64-dimensional embedding, selected from a dimensionality sweep reported in Appendix~\ref{app:dimension}. It contains 177\,344 trainable parameters, approximately equally split between the light curve encoder and the periodogram encoder. We trained it for 20 epochs. The full training configuration, hardware details, and the experimental justification for the architectural and VICReg hyperparameter choices are all presented in Appendix~\ref{app:model_development}.

\section{Results}\label{sec:results}

We now evaluate the representations produced by the framework described in the preceding sections. We assess whether the frozen embeddings capture known variability classes, examine the contributions of the individual input modalities, analyse the continuous physical structure encoded in the embedding space, and describe the publicly released code.

\subsection{Classification performance}\label{sec:res_classification}

Table~\ref{tab:classification} reports the test-set performance of the LR and SVM classifiers trained on the frozen embeddings, evaluated at the three hierarchy levels defined in Sect.~\ref{sec:catalogue}. At the full 14-class level, the LR achieves 77.1\% balanced accuracy and the SVM 78.4\%, a gap of 1.3 percentage points. This narrow gap indicates that the variability classes are already nearly linearly separable in the embedding space, and that a simple linear classifier captures most of the available class structure.

Performance increases substantially at the 12-class level, where AGN, QSO, and Blazar are merged into a single class, reaching 84.3\% (LR) and 85.1\% (SVM). The improvement reflects the strong mutual confusion among these three classes at the 14-class level, which share similar stochastic variability patterns. At the broadest level of four physical categories, the LR reaches 90.5\% and the SVM 92.6\%.

\begin{table}[!ht]
\caption{Test-set classification performance.}
\label{tab:classification}
\centering
\begin{tabular}{l c c c c c}
\hline\hline
 & & \multicolumn{2}{c}{LR} & \multicolumn{2}{c}{SVM} \\
\cline{3-4}\cline{5-6}
Level & $N$ & BAcc & F1 & BAcc & F1 \\
\hline
Full    & 14 & 77.1 & 72.4 & 78.4 & 74.2 \\
Grouped & 12 & 84.3 & 80.4 & 85.1 & 81.6 \\
Broad   &  4 & 90.5 & 84.6 & 92.6 & 87.7 \\
\hline
\end{tabular}
\tablefoot{Balanced accuracy (BAcc) and macro F1, per cent, for LR and a radial-basis SVM, evaluated on the held-out test set.}
\end{table}

\begin{figure}[!ht]
  \centering
  \includegraphics[width=\columnwidth]{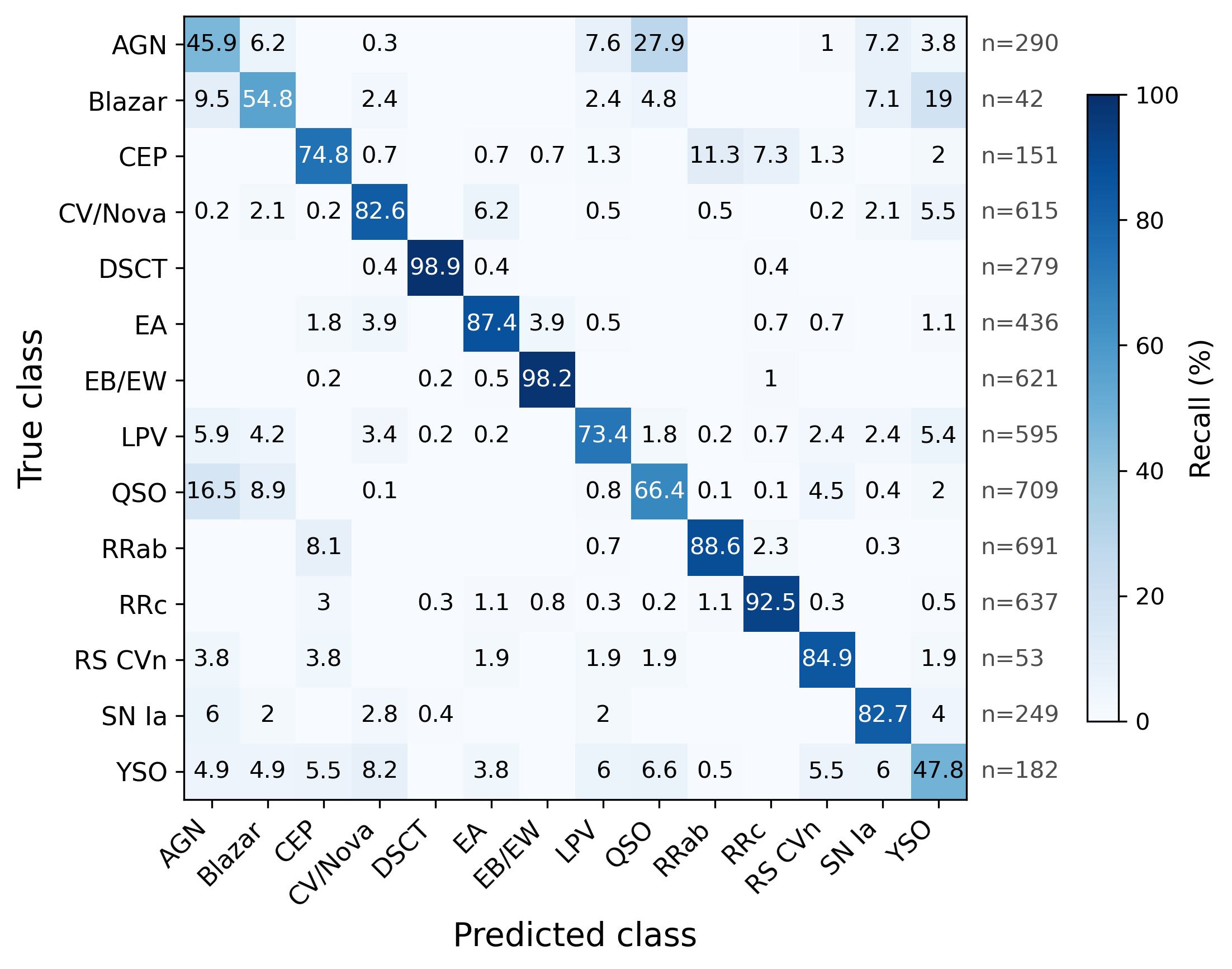}
  \includegraphics[width=\columnwidth]{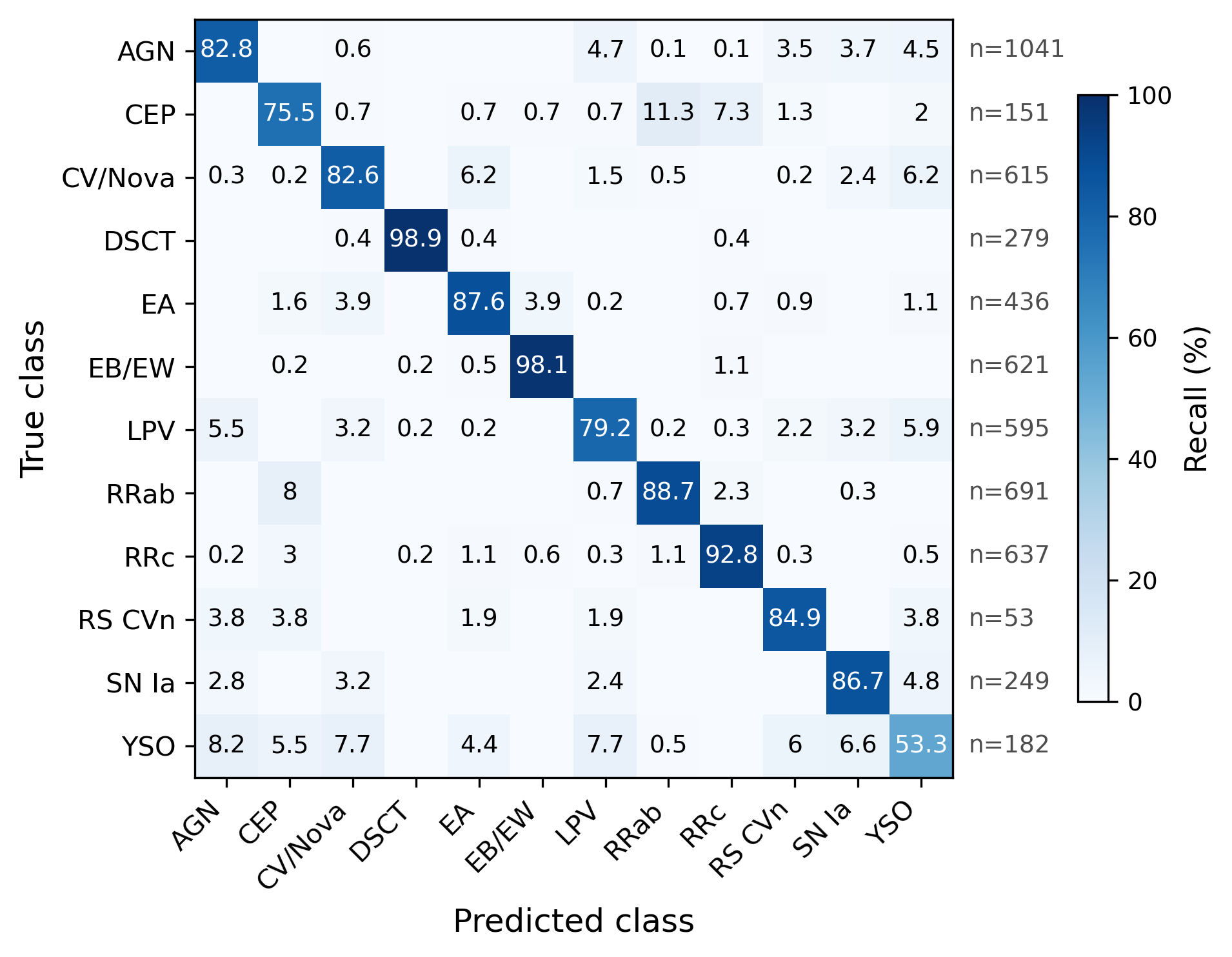}
  \caption{Logistic regression confusion matrices on the test set. \textit{Top:} 14 individual classes. \textit{Bottom:} 12 classes with AGN, QSO, and Blazar merged. Rows are true classes, columns are predictions, and cell values give the percentage of sources in each true class assigned to each predicted class. Row sample sizes are indicated on the right.}
  \label{fig:confusion_matrices}
\end{figure}

Figure~\ref{fig:confusion_matrices} shows the LR confusion matrices at the 14-class and 12-class levels. At the 14-class level, $\delta$~Scuti variables (98.9\% recall) and contact/semi-detached eclipsing binaries (98.2\%) are classified near-perfectly. RR~Lyrae first-overtone variables (92.5\%), RR~Lyrae fundamental-mode variables (88.6\%), detached eclipsing binaries (87.4\%), RS~CVn variables (84.9\%), cataclysmic variables (82.6\%), and Type~Ia supernovae (82.7\%) all exceed 80\% recall. Cepheids (74.8\%) show confusion primarily with RR~Lyrae variables, and long-period variables (73.4\%) are scattered across several classes. YSOs achieve the lowest recall (47.8\%), with misclassifications distributed broadly across the full class set, a consequence of their extreme morphological diversity discussed in Sect.~\ref{sec:disc_yso}. At the 12-class level, the merged AGN class reaches 82.8\% recall, a large improvement over the individual subclasses. Most other classes show modest gains, though YSO remains the most challenging class at 53.3\%.

\subsection{Input modality comparison}\label{sec:res_modality}

Table~\ref{tab:modality} compares the classification performance when the model is trained with both input modalities, with the light curve branch only, or with the periodogram branch only. These are independently trained models evaluated through 5-fold CV on the training and validation set, as described in Sect.~\ref{sec:eval_labels}. The combined model achieves 76.8\% balanced accuracy (LR) at the 14-class level, substantially outperforming both the light curve-only model (62.6\%) and the periodogram-only model (53.8\%). Neither single modality approaches the combined performance, confirming that the branches capture complementary information.

\begin{table}[!ht]
\caption{Input modality comparison.}
\label{tab:modality}
\centering
\begin{tabular}{l c c c}
\hline\hline
Input & LR & SVM & SVM $-$ LR \\
\hline
Light curve only & 62.6 & 60.4 & $-2.2$ \\
Periodogram only  & 53.8 & 64.3 & $+10.5$ \\
Combined          & 76.8 & 77.3 & $+0.5$ \\
\hline
\end{tabular}
\tablefoot{Balanced accuracy at the 14-class level, per cent, from five-fold cross-validation on the training and validation set.}
\end{table}

The SVM$-$LR gap reveals the geometry of each embedding. The combined embedding shows a gap of only $+0.5$ points, confirming near-linear separability. The light curve-only embedding shows a gap of $-2.2$ points, where the SVM performs slightly worse than the LR, suggesting that the light curve representation is already linearly separable and the additional flexibility of the SVM leads to slight overfitting. The periodogram-only embedding shows the largest gap of $+10.5$ points, indicating that the periodogram alone produces a representation with highly non-linear class boundaries. Combining both modalities resolves this non-linearity into a linearly separable representation.

The improvement from combining modalities is not uniform across classes. Geometric variables such as detached and contact eclipsing binaries achieve high recall with the light curve branch alone (90.4\% and 95.7\%, respectively), as their distinctive eclipse shapes are directly visible in the time-domain data. Pulsating variables benefit strongly from the periodogram branch, with Cepheids improving from 38.2\% (LC-only) to 75.1\% (combined) and fundamental-mode RR~Lyrae from 81.9\% to 87.6\%. Long-period variables show the largest combined gain, improving from 16.6\% (LC-only) and 64.5\% (PG-only) to 73.3\% (combined), demonstrating that both modalities contribute substantial and complementary information for this class. The per-class recall for all three configurations is reported in Appendix~\ref{app:modality_perclass}.

\subsection{Embedding structure and light curve properties}\label{sec:res_properties}

Table~\ref{tab:correlations} reports the Spearman rank correlation between eight properties and the embedding dimension with which each correlates most strongly. Variability amplitude shows the highest correlation ($|\rho| = 0.804$), indicating that the embedding strongly encodes the strength of the brightness variations. The light curve asymmetry index $M$ is similarly well captured ($|\rho| = 0.788$), as is the skewness of the magnitude distribution ($|\rho| = 0.775$), which maps to a different embedding dimension and therefore captures a complementary aspect of the light curve shape. The quartile ratio $Q_{31}$, a robust measure of the magnitude spread, correlates at $|\rho| = 0.608$. The periodicity index $Q$ shows a correlation of $|\rho| = 0.586$, reflecting the separation between periodic and aperiodic sources. The autocorrelation length, which measures the characteristic timescale of the variability, correlates at $|\rho| = 0.581$ on a different dimension from the GLS period ($|\rho| = 0.480$), indicating that the embedding separately encodes the dominant period and the overall timescale structure. The GLS period is expected to show a moderate rather than strong correlation because it is computed for all sources including stochastic variables, where the dominant peak is typically a spurious alias. A strong correlation with period would indicate overfitting to a quantity that is astrophysically meaningless for a large fraction of the dataset. The number of observations per source shows the weakest correlation ($|\rho| = 0.327$), confirming that observation count is not a dominant axis of the embedding. Each of the eight properties maps to a different embedding dimension, indicating that the representation distributes distinct physical information across separate dimensions rather than concentrating it. An extended correlation analysis with additional features is presented in Appendix~\ref{app:correlations}.

\begin{table}[!ht]
\caption{Light curve property correlations with the embedding.}
\label{tab:correlations}
\centering
\begin{tabular}{l c c}
\hline\hline
Property & Dim. & $|\rho|$ \\
\hline
Amplitude              & 39 & 0.804 \\
$M$                    & 29 & 0.788 \\
Skewness               & 53 & 0.775 \\
$Q_{31}$               &  8 & 0.608 \\
$Q$                    & 34 & 0.586 \\
Autocorrelation length & 43 & 0.581 \\
GLS period             & 22 & 0.480 \\
$n_\mathrm{final}$     & 30 & 0.327 \\
\hline
\end{tabular}
\tablefoot{Spearman rank correlation between each property and the most strongly correlated embedding dimension, computed on the full dataset.}
\end{table}

\begin{figure}[!h]
  \centering
  \includegraphics[width=\columnwidth]{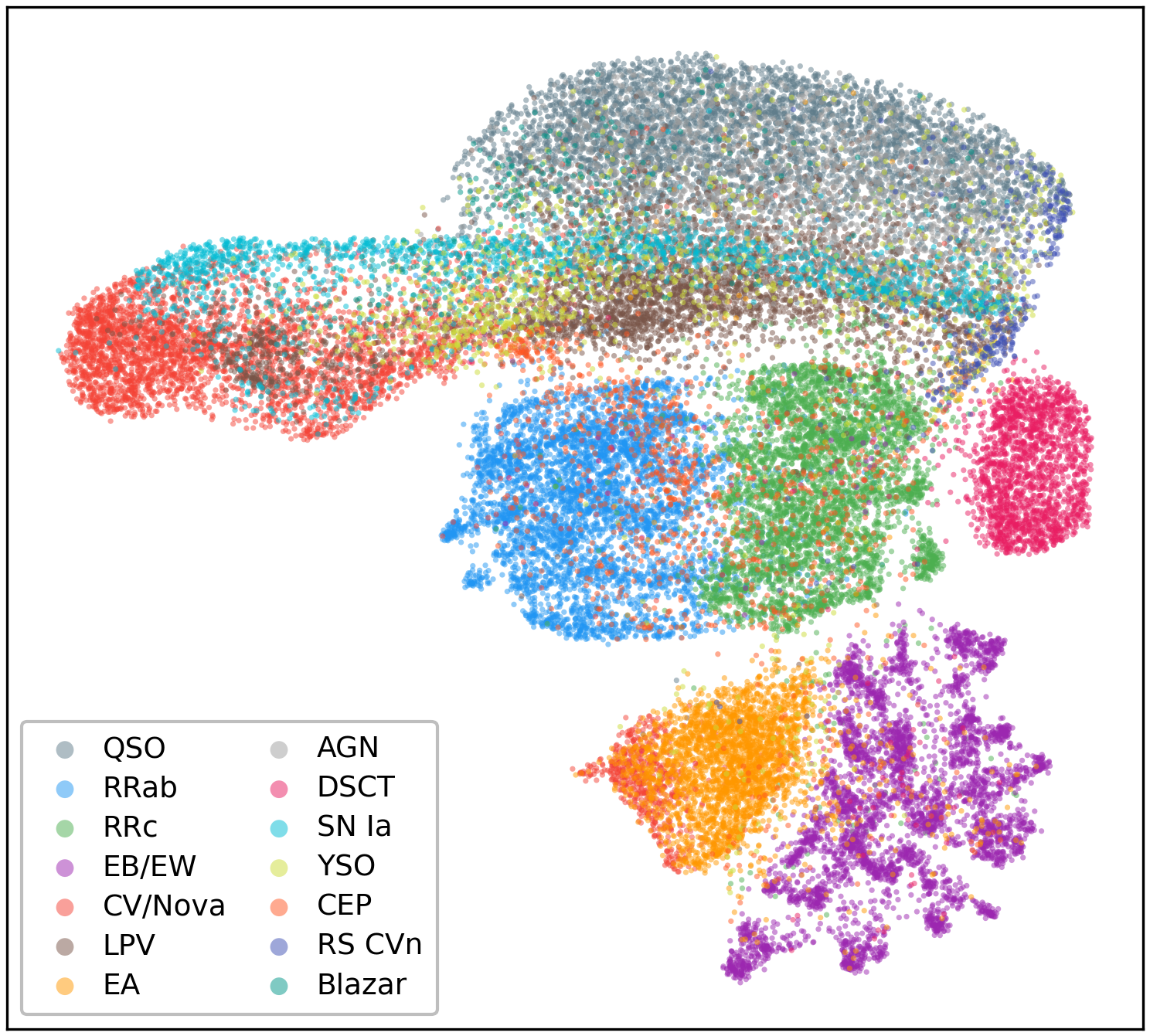}
  \caption{Two-dimensional UMAP projection of the 64-dimensional embedding, computed on the full dataset and coloured by the 14 variability classes. UMAP projections depend on the choice of hyperparameters: this projection uses \texttt{n\_neighbors} $=100$ and \texttt{min\_dist} $=0.5$.}
  \label{fig:umap_class}
\end{figure}

Figure~\ref{fig:umap_class} shows a 2D UMAP projection of the 64D embedding, coloured by the 14 variability classes. The EB/EW forms a distinct cluster at the bottom, fragmented into subclusters. EAs sit nearby, together with a small subset of cataclysmic variables that are eclipsing binaries (Sect.~\ref{sec:disc_confusions}). The pulsating variables occupy the middle region, arranged from left to right as fundamental-mode RR~Lyrae, first-overtone RR~Lyrae, and $\delta$~Scuti, with Cepheids positioned between the two RR~Lyrae groups. Above the pulsating region, Type~Ia supernovae, LPVs, CVs, and YSOs form a filamentary structure that overlaps in this two-dimensional projection (but separates more clearly in three). The AGN, QSO, and Blazar classes occupy the uppermost region with substantial mutual overlap, consistent with their shared physical origin. RS~CVn variables sit at the right edge.

Figures~\ref{fig:umap_amplitude}--\ref{fig:umap_skewness} in Appendix~\ref{app:correlations} show the same projection coloured by four light curve properties. The amplitude colouring reveals a continuous left-to-right gradient from high-amplitude to low-amplitude sources. The periodicity colouring shows a vertical arrangement from periodic sources with low $Q$ at the bottom and middle, where the EB/EW, RR~Lyrae, and $\delta$~Scuti clusters are located, to fully aperiodic sources with $Q \approx 1$ at the top. The asymmetry index $M$ \citep{Cody2014} and skewness vary along different axes, confirming that these two measures capture complementary information encoded in separate embedding dimensions.

\subsection{Multi-band evaluation}\label{sec:res_multiband}

The architecture presented in Sect.~\ref{sec:architecture} accepts single-band photometric time series as input. Extending the framework to process multi-band light curves jointly is a planned direction for future work. Nevertheless, we can train separate models on different photometric bands independently and examine whether their learned representations capture complementary information.

We trained two additional models from scratch on ZTF $g$-band and $i$-band light curves, using the same architecture, hyperparameters, and training procedure as the $r$-band model described in the preceding sections. The only difference between the three models is the input data, as each uses the single-band light curves available for its respective filter from the same ALeRCE source catalogue. After preprocessing, 35\,737 sources remained in $g$ and 17\,978 in $i$, compared to 37\,006 in $r$ (Table~\ref{tab:catalogue}).

To ensure a fair comparison across bands, we restricted the evaluation to the 17\,442 sources that are present in all three preprocessed datasets, limited primarily by the $i$-band coverage. For each of these common sources, we computed the frozen 64D embedding from each of the three independently trained models. We then evaluated not only each single-band embedding on its own, but also all pairwise and the triple-band combinations. The combined embeddings were formed by concatenating the independent frozen vectors (producing 128D vectors for pairs and a 192D vector for the triple combination) and training an LR classifier on the embeddings through 5-fold stratified CV.

Table~\ref{tab:multiband} reports the balanced accuracy for all seven configurations at the three class hierarchy levels. The three single-band models achieve similar performance at the 14-class level (75.4\% for $r$, 76.1\% for $g$, 73.8\% for $i$). Every pairwise combination outperforms every single band, and the best pair ($g$+$i$, 84.8\%) outperforms the other two pairs by 2.5 percentage points despite identical dimensionality. The full three-band concatenation reaches 86.1\% balanced accuracy at the 14-class level, 91.4\% at the 12-class level, and 96.0\% at the broad category level, improving over the best single band by 10.0, 7.9, and 3.4 percentage points, respectively. Per-class recall for all configurations is reported in Appendix~\ref{app:multiband}.

\begin{table}[!ht]
\caption{Multi-band embedding evaluation.}
\label{tab:multiband}
\centering
\begin{tabular}{l r c c c}
\hline\hline
Embedding & Dim & class (14) & mid (12) & broad (4) \\
\hline
$g$          &  64 & 76.1 & 83.5 & 92.6 \\
$r$          &  64 & 75.4 & 82.1 & 91.8 \\
$i$          &  64 & 73.8 & 81.2 & 87.7 \\
$g$+$r$      & 128 & 82.3 & 88.2 & 94.5 \\
$g$+$i$      & 128 & 84.8 & 90.5 & 95.4 \\
$r$+$i$      & 128 & 82.1 & 89.2 & 94.2 \\
$g$+$r$+$i$  & 192 & 86.1 & 91.4 & 96.0 \\
\hline
\end{tabular}
\tablefoot{Balanced accuracy (per cent) from 5-fold CV with an LR classifier on the 17\,442 sources common to all three bands.}
\end{table}

\subsection{Public code and framework}\label{sec:res_code}

The trained model, source code, and a demonstration dataset with tutorials are publicly released\footnote{\url{https://github.com/xMediMadix/multimodal-varstar}}. The framework is designed to be survey-agnostic and requires no code modifications to apply to different photometric surveys or label sets. All data-specific choices, including the survey band, preprocessing strategy, period search range, standardisation mode, and augmentation parameters, are controlled through configuration files. Users provide their own light curve data and any set of class labels or continuous properties for evaluation, and the framework produces embeddings, trains the evaluation classifiers, and generates diagnostic plots automatically.

The method is not limited to broad variability classification across many classes. It can equally be applied within a single variability type to discover substructure, identify outliers, or reveal morphological diversity that is not captured by existing taxonomies. We encourage scientists to explore such applications within their own domain of expertise, using the config-driven framework with their own catalogues. The demo dataset with usage instructions is included in the repository to facilitate getting started. The second paper in this series will demonstrate such a focused application to young stellar objects.

\section{Discussion}\label{sec:discussion}

\subsection{Class confusions}\label{sec:disc_confusions}

The confusion matrices in Fig.~\ref{fig:confusion_matrices} show that most misclassifications follow patterns that are astrophysically expected given the light curve morphologies involved.

\begin{figure}[!ht]
  \centering
  \includegraphics[width=\columnwidth]{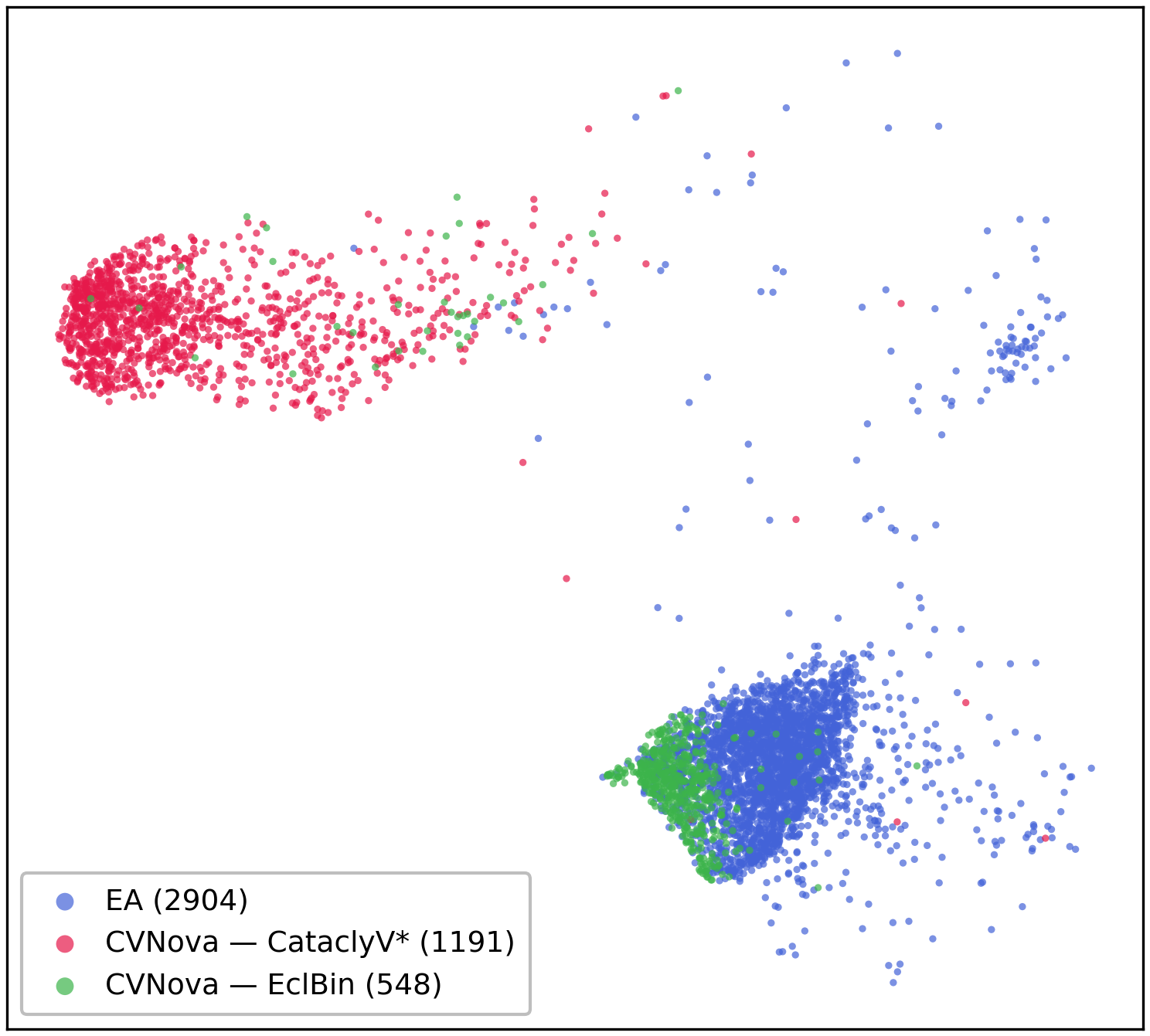}
  \caption{CV/Nova and EA populations in the UMAP projection. EA sources are shown in blue (2904), CV/Nova sources classified as CataclyV* by SIMBAD in red (1191), and CV/Nova sources classified as EclBin by SIMBAD in green (548).}
  \label{fig:cvnova_ea}
\end{figure}

The most informative case concerns the confusion between cataclysmic variables and detached eclipsing binaries. The confusion matrix shows that 6.2\% of CV/Nova sources are assigned to the EA class by the linear classifier, and 3.9\% of EA sources are assigned to CV/Nova. To investigate this, we cross-matched all 4098 ALeRCE CV/Nova sources with SIMBAD, finding that 1191 (29\%) are confirmed cataclysmic variables (CataclyV*) while 548 (13\%) are classified as eclipsing binaries (EclBin). Figure~\ref{fig:cvnova_ea} shows the EA population together with the CV/Nova sources split by their SIMBAD classification. The confirmed cataclysmic variables (red) form a compact group in the upper left of the projection, well separated from the EA cluster (blue) in the lower right. The 548 CV/Nova sources that SIMBAD classifies as eclipsing binaries (green), however, sit directly on top of the EA cluster, showing that their light curve morphology is dominated by eclipses. The embedding has placed these sources near the eclipsing binaries based on light curve similarity alone, reflecting the dominant variability signal regardless of the ALeRCE label. This provides an independent confirmation that class labels did not influence the learned representation. If label information had leaked into the training objective, these sources would have remained within the CV/Nova region regardless of their true light curve morphology.

The Cepheid class shows a different pattern. Cross-matching the 1005 ALeRCE Cepheids with SIMBAD reveals a heterogeneous composition of 57\% classical Cepheids, 13\% Type~II Cepheids, 6\% eclipsing binaries, 6\% RS~CVn variables, and 6\% RR~Lyrae stars. The eclipsing binaries and RR~Lyrae found among the ALeRCE Cepheids are placed in the EA and RR~Lyrae regions of the embedding, respectively, again indicating that the embedding organises these sources by their dominant light curve morphology rather than by the ALeRCE label. The residual confusion between genuine Cepheids and RR~Lyrae variables in the classification is expected, as both are radially pulsating stars occupying the classical instability strip with similar waveform shapes and partially overlapping period ranges \citep{bhardwaj2020}.

The AGN, QSO, and Blazar classes show strong mutual confusion at the 14-class level, which is largely resolved by merging them at the 12-class level as discussed in Sect.~\ref{sec:res_classification}. Within the framework of AGN unification \citep{Padovani2017}, these classes are understood as different observational manifestations of the same underlying process of accretion onto a supermassive black hole, with their observed properties largely set by orientation and jet-related effects. Their optical variability is well described by stochastic processes associated with the accretion disk \citep{MacLeod2010}, so separating them reliably from single-band photometry alone is not expected.

\subsection{Within-class structure}\label{sec:disc_structure}

Supervised classifiers are limited to the classes defined by their training labels and cannot reveal finer structure within those classes. Unsupervised and self-supervised embeddings are not subject to this constraint, and the representations learned by our framework show astrophysically meaningful substructure within several of the broad ALeRCE classes.

The long-period variable class provides the clearest example. Although all LPV sources are placed in a single coherent region in the full class UMAP of Fig.~\ref{fig:umap_class}, a closer inspection reveals two distinct groups, as shown in Fig.~\ref{fig:umap_lpv} in Appendix~\ref{app:disc_figures}. Cross-matching all 3966 ALeRCE LPV sources with SIMBAD yields 63\% classified as LongPeriodV*, 22\% as LongPeriodV* candidates, and 9.5\% as Mira variables. The Mira variables correspond to the smaller, more compact group separated from the larger diffuse population. Mira variables are distinguished from other LPV subtypes by their large visual pulsation amplitudes and relatively regular, single-mode pulsation, which produces much more stable light curves than in semi-regular variables \citep{WoodSebo1996}. Within the larger diffuse group, amplitude is the dominant gradient, with a weaker but visible gradient in periodicity also visible in Fig.~\ref{fig:umap_lpv}.

The contact and semi-detached eclipsing binary class (EB/EW) shows a different form of internal organisation, with the UMAP cluster fragmenting into distinct subclusters corresponding to narrow ranges of orbital periods, as shown in Fig.~\ref{fig:umap_ebew} in Appendix~\ref{app:disc_figures}. The model organises eclipsing binaries by period within the class, even though period is never used explicitly as a training target and enters the representation only through the periodogram branch.

These findings demonstrate that the framework is not limited to broad variability classification but can equally serve as a tool for within-class analysis, revealing substructure that reflects genuine astrophysical differences among sources sharing the same broad classification.

\subsection{YSO morphological diversity}\label{sec:disc_yso}

YSOs achieve the lowest recall of all 14 classes (47.8\%), with misclassifications distributed broadly across the full class set rather than toward any single class. This is not necessarily a failure of the model but rather a consequence of the extreme morphological diversity of YSO light curves \citep{Cody2014, Mas2026}.

To illustrate this, we trained the framework from scratch on a subset of the NGYSO catalogue \citep{Marton2026_ngyso}, selecting of approximately 27\,000 YSO candidates observed by ZTF. We assigned morphological types to each source by thresholding the periodicity index $Q$ and asymmetry index $M$ of \citet{Cody2014}, as demonstrated at scale on Gaia DR3 light curves by \citet{Mas2026}. The resulting labels are noisy, as the hard thresholds impose discrete boundaries on what is a continuous distribution in the $Q$--$M$ plane, but they are sufficient for this demonstration. Figure~\ref{fig:umap_yso} in Appendix~\ref{app:disc_figures} shows the UMAP projection of the resulting embeddings, coloured by the five main morphological types used in this experiment. Periodic sources (P), stochastic variables (S), bursters (B), aperiodic dippers (APD), and quasi-periodic dippers (QPD) each occupy distinct regions of the embedding. Bursters and dippers (APD and QPD combined) are located on opposite sides of the projection, reflecting their contrasting light curve behaviour and opposite signs of $M$ \citep{Cody2014}. A continuous gradient runs from P sources through QPDs to APDs, tracing the transition from regular to irregular variability. Stochastic variables fill the intermediate space between the bursters and the dippers.

This morphological diversity explains the low classification accuracy on the ALeRCE YSO sample. Periodic YSOs produce light curves resembling rotational variables, bursters resemble cataclysmic variables, and aperiodic dippers can look similar to stochastic AGN variability. The embedding correctly places each YSO near the class whose light curve it most closely resembles, but this distributes the YSO population across the embedding and prevents any single linear boundary from separating them from all other classes. The authors will present a dedicated in-depth analysis of YSO variability morphology using this framework in an upcoming study.

\subsection{Comparison with reconstruction-based approaches and limitations}\label{sec:disc_comparison}

Reconstruction-based approaches such as variational autoencoders \citep{Kingma2014, Huijse2025} learn representations by training a decoder to reproduce the input light curve point by point, creating several difficulties that the presented contrastive approach avoids.

Consider two sources with the same type of brightening event, such as an accretion burst or a supernova explosion, occurring at different times. A VAE must encode the event timing to reconstruct each light curve correctly, which may place them in different regions of the latent space despite identical variability behaviour. Our split-half augmentation produces views where the event observations are distributed differently between the two halves, from roughly even splits to cases where one half captures most of the event while the other contains little or none of it. This encourages the model to extract global event properties such as amplitude, duration, and rise profile rather than absolute timing. The clean separation of Type~Ia supernovae (86.7\% recall; see bottom panel of Fig.~\ref{fig:confusion_matrices}), a class defined by non-repeating transient events, is consistent with this expected advantage.

Consider next a detached eclipsing binary observed by a ground-based survey such as ZTF. The vast majority of observations fall on the out-of-eclipse baseline, while each individual eclipse is typically sampled by only a handful of points. A VAE reconstruction loss averaged over all observations is dominated by this flat baseline, learning to reproduce the non-fading level while smoothing over the sparsely sampled eclipses. Phase-folding the light curve before encoding, as done by \citet{Huijse2025}, can mitigate this for periodic sources, but it excludes aperiodic and stochastic variables by construction. The contrastive objective used in this study faces no such pressure, since both halves share the same eclipse statistics regardless of which specific events each half contains, and no folding is required. The high 87.6\% recall of detached eclipsing binaries (bottom panel of Fig.~\ref{fig:confusion_matrices}) is consistent with this advantage.

When the dataset mixes low-amplitude and high-amplitude variables, the VAE reconstruction loss can be dominated by the majority of low-amplitude sources, making it difficult to accurately reconstruct the rarer high-amplitude variables whose residuals are disproportionately large relative to the bulk of the training data. The contrastive objective used here is not affected by this because the invariance loss compares each source only with itself, and the global normalisation described in Sect.~\ref{sec:normalisation} preserves relative amplitude differences without favouring any amplitude range. The strong encoding of amplitude across the full dynamic range ($|\rho| = 0.804$, Table~\ref{tab:correlations}), visible as a smooth gradient in Fig.~\ref{fig:umap_amplitude}, confirms that sources of all amplitudes are represented uniformly in the embedding.

On the other hand, some limitations of the present approach and study should be noted. The split-half augmentation and length augmentation that underpin the training strategy naturally favour surveys with long light curves. Surveys with fewer data points per source on average, such as Gaia DR3, would provide less diverse temporal subsets for each half and are therefore expected to yield weaker representations. For surveys with long light curves, however, this property becomes an advantage over reconstruction-based approaches, which may struggle to accurately reproduce every observation in a long and noisy sequence.

The evaluation relies on ALeRCE labels that are themselves imperfect. The classification accuracies reported in Sect.~\ref{sec:res_classification} should be interpreted with this in mind.

Although the framework is designed to operate on single-band photometry, Sect.~\ref{sec:res_multiband} demonstrates that independently trained single-band models capture complementary wavelength-dependent information that can be combined post-hoc by concatenating their frozen embeddings. This multi-band extension requires no architectural modification or joint training. The released code can also be extended to accept multi-band input directly by treating colour information as an additional modality, which we plan to explore in the upcoming YSO study.

\subsection{Multi-band complementarity}\label{sec:disc_multiband}

The multi-band evaluation of Sect.~\ref{sec:res_multiband} reveals several findings worth discussing. The $i$-band model achieves 73.8\% balanced accuracy at the 14-class level despite operating on roughly half the training data (17\,978 sources compared to 37\,006 in $r$) and a median of only 152 observations per source, roughly a quarter of the $r$-band median of 679. This demonstrates that the framework does not require hundreds of observations per source to learn meaningful representations. The split-half augmentation remains effective at these observation counts.

The concatenation results show that each independently trained single-band model encodes different physical aspects of the variability, because different wavelengths probe different astrophysical processes. The $g$+$i$ combination (84.8\%) outperforms the other pairs ($r$+$g$ at 82.3\% and $r$+$i$ at 82.1\%) despite identical dimensionality, which argues against a pure dimensionality effect and points to wavelength-dependent information. The $g$ and $i$ bands span the widest wavelength baseline ($\sim$480 to $\sim$790~nm), maximising the colour leverage available from ZTF photometry. The dimensionality sweep of Appendix~\ref{app:dimension} provides further evidence for this interpretation. A single $r$-band model trained with 128 embedding dimensions achieves only 77.4\% (Table~\ref{tab:dimension}), gaining just 1.2 percentage points from doubling the dimensionality relative to the 64-dimensional baseline. By contrast, the 128-dimensional concatenated pairs gain 6.9 to 9.4 percentage points over the best single band, despite the fact that the two constituent embeddings were never jointly optimised. The improvement therefore reflects complementary wavelength-dependent information rather than the additional capacity of a higher-dimensional embedding.

The per-class results in Table~\ref{tab:multiband_perclass} show that the majority of classes, including pulsating stars \citep[consistent with pulsation amplitudes being largest at shorter wavelengths,][]{Varma2024, Mowlavi2021}, eclipsing binaries (EA, EB/EW), and QSOs, achieve their highest single-band recall in the $g$ band. A smaller group of intrinsically red or embedded sources, LPVs, AGN, and YSOs, perform best in the $i$ band. Across all classes, every pairwise and triple-band concatenation outperforms every single band at the aggregate level, confirming that independently trained models capture complementary information.

The wavelength-dependent amplitude ratios of different variability types are a known diagnostic tool in time-domain astrophysics. For example, \citet{Mowlavi2021} showed that Gaia multi-band amplitude proxies can categorise variable star types into four groups according to their blue-to-red amplitude ratios and colours. Our concatenation experiment arrives at a complementary result from the opposite direction. Rather than computing explicit amplitude ratios, each single-band model learns its own variability representation, and colour information emerges implicitly from their concatenation.

The present study was developed and evaluated on ZTF, which serves as a precursor to LSST \citep{Ivezic2019}. LSST will survey the southern sky in six filters ($ugrizy$) that include the same $g$, $r$, and $i$ bands used here, delivering long-baseline, densely sampled multi-band light curves over ten years \citep{Ivezic2019, Hambleton2023}. The results of this study suggest that the framework is well suited to these conditions, as we demonstrated that it learns meaningful representations across a wide range of variable star types and benefits from multi-band information.

\section{Summary}\label{sec:conclusions}

We have presented a self-supervised framework for the characterisation of variable source light curves that operates entirely without labelled training data. The method combines two complementary views of each source, raw time-domain observations processed by a Transformer encoder and a GLS periodogram processed by a ResNet, trained jointly with the VICReg objective on split-half temporal augmentations. The framework produces compact 64-dimensional embeddings that capture light curve properties of stellar variability from the observations.

Evaluated on 37\,006 ZTF sources spanning 14 variable source classes, simple linear classifiers trained on the frozen embeddings achieve 77.1\% balanced accuracy at the individual class level and 90.5\% at the broad category level, with a narrow LR--SVM gap confirming near-linear separability. Combining the light curve and periodogram branches improves balanced accuracy by up to 14.9 percentage points over the best single modality for individual classes, demonstrating that the two branches capture complementary information. The embedding encodes a range of light curve properties including variability amplitude, light curve asymmetry, skewness, periodicity, and characteristic timescale, each mapped to a distinct embedding dimension.

The learned representations reveal structure that goes beyond the classification labels. Cross-matching with SIMBAD shows that 99.4\% of the CV/Nova sources placed in the eclipsing binary region of the embedding are confirmed eclipsing binaries. Within-class analysis of long-period variables recovers the separation of Mira variables from semi-regular and irregular subtypes, and the EB/EW population fragments into period-defined subclusters. The low recall of young stellar objects (47.8\%) is explained by their extreme morphological diversity, as demonstrated by training the framework on a dedicated YSO catalogue where bursters, dippers, periodic, and stochastic sources occupy distinct embedding regions.

Retraining the identical architecture on ZTF $g$-band and $i$-band data without modification reproduces the main results, with the $i$-band model achieving 73.8\% balanced accuracy despite roughly half the training data and four times fewer observations per source. Concatenating the frozen embeddings from all three bands reaches 86.1\% at the 14-class level, an improvement of 10 percentage points over the best single band, demonstrating that independently trained models capture complementary wavelength-dependent information. The framework is directly applicable to LSST, which will provide multi-band time series in the same filters with long, densely sampled light curves.

The trained model and the source code with tutorials are publicly available as described in Sect.~\ref{sec:res_code}. The framework is survey-agnostic and config-driven, requiring no code modifications to apply to different photometric surveys, label sets, or science cases. The authors will present an in-depth application to young stellar object variability in an upcoming study.

\begin{acknowledgements}
This research has made use of the SIMBAD database and the VizieR catalogue access tool, operated at CDS, Strasbourg, France \citep{Wenger2000}. This research has made use of the cross-match service provided by CDS, Strasbourg. This research has made use of the NASA/IPAC Infrared Science Archive, which is funded by the National Aeronautics and Space Administration and operated by the California Institute of Technology. M.M. and G.M. acknowledge support from the ESA PRODEX contract nr. 4000132054. M.A. acknowledges support by the Swiss National Science Foundation, grant 2000-1-240076.

\end{acknowledgements}

%
   \bibliographystyle{aa} 
   \bibliography{ref} 
\begin{appendix} 
\section{Model development and implementation}\label{app:model_development}

\subsection{Architecture experiments}\label{app:architecture_experiments}

The light curve Transformer and periodogram ResNet, introduced in Sects.~\ref{sec:lc_encoder} and~\ref{sec:pg_encoder}, each produce a fixed-size output vector of $d_\mathrm{LC}$ and $d_\mathrm{PG}$ dimensions, respectively. Before combining them into a single embedding, each branch output can optionally pass through a learned non-linear projection that maps it to a different dimensionality $d_\mathrm{proj}$. For example, a 128-dimensional encoder output could be projected to 64 dimensions through a linear layer followed by a GELU activation, reducing the number of off-diagonal elements that the VICReg covariance term must decorrelate from $128 \times 127 = 16\,256$ to $64 \times 63 = 4\,032$. After projection, the two branch vectors are concatenated and passed through a fusion layer, a small network that maps the concatenated vector to the final embedding dimension $d_\mathrm{embed}$. We explored varying the projection dimensionality, including configurations where the encoder output was compressed to a lower-dimensional representation before fusion, as well as deeper fusion layers and non-linear activations.

When non-linear projection layers were used to compress the branch outputs, or when the fusion layer contained additional hidden layers, the training loss decreased rapidly in the first few epochs. However, the validation loss diverged shortly after, producing a large and growing gap between training and validation loss. This pattern indicates that the additional learnable parameters allowed the model to memorise source-specific features from the training set rather than learning general variability patterns. The overfitting also set in significantly earlier compared to configurations without learned projections, leaving less training time for the model to develop refined representations.

By contrast, when both branch outputs were passed directly to the fusion layer without intermediate projection and the fusion layer consisted of a single linear layer followed by a GELU activation, training remained stable across all 20 epochs without an increase in validation loss. The single-layer design forces each encoder branch to produce representations that are already informative, so the fusion layer only needs to recombine them in one step rather than learning deeper cross-modal interactions.

The final architecture uses both encoders with a 64-dimensional output, identity projections (no learned transformation), and a single fusion layer that maps the 128-dimensional concatenation to the 64-dimensional embedding. The full specification is listed in Table~\ref{tab:architecture}.

\subsection{VICReg hyperparameter analysis}\label{app:vicreg_hyperparameters}

The VICReg loss weights and the variance target $\gamma$ were selected through a systematic sweep during label-free development, using the validation loss and UMAP projections as selection criteria.

The invariance weight $\lambda_\mathrm{inv}$ was reduced from the default value of 25 used in the original computer vision application \citep{Bardes2022} to 5. This adjustment was motivated by the periodogram branch. For periodic sources, the two split-half periodograms share the same dominant peaks, and a strong invariance pressure correctly encourages matching representations. For stochastic and aperiodic sources, however, the periodograms computed on each half are dominated by noise and their peaks are spurious. A high invariance weight penalises the model for failing to match these inherently dissimilar periodograms, even though matching them carries no astrophysical meaning. With $\lambda_\mathrm{inv} = 5$, periodic halves are still pulled together effectively, but the reduced pressure allows the model to instead learn that flat, noise-dominated periodograms are themselves a characteristic signature that separates non-periodic from periodic sources.

The covariance weight $\lambda_\mathrm{cov}$ was set to 5. At higher values, we observed a training pattern similar to the projection experiments described above. The model reached its best validation loss within a few epochs while the training loss continued to decrease, indicating early overfitting. With $\lambda_\mathrm{cov} = 5$, training proceeded stably over all 20 epochs, allowing the model to develop more refined features. The stronger covariance pressure appears to push the embedding dimensions toward statistically independent directions too aggressively, producing features that decorrelate well on the training batch but fail to generalise.

The variance target was set to $\gamma = 0.2$, lower than the default of 1.0. This produced both improved training stability and better-structured UMAP projections. With larger $\gamma$ values, the UMAP projections showed many small, fragmented clusters, suggesting that the high variance floor forced the embedding to spread into non-linear directions. With $\gamma = 0.2$, the projections showed more coherent clusters indicating better linear separability.

We apply the VICReg loss in combined mode, computing it on the fused embedding and independently on each branch embedding. We compared this against two alternatives. In the first, no fusion layer was used and the two 64-dimensional branch outputs were simply concatenated into a 128-dimensional vector, with the VICReg loss computed on this concatenation. Without a layer that can combine features across the two modalities, the embedding showed poor dimension health with several collapsing dimensions, and the model could not learn cross-modal relationships between the time-domain and frequency-domain representations. In the second alternative, a fusion layer was present but the VICReg loss was computed only on the fused output. Because the individual branches received no direct variance or covariance pressure, some branch dimensions collapsed before reaching the fusion layer, reducing the effective information available to the final embedding. The combined mode avoids both failure cases. Each branch independently maintains well-conditioned representations through its own loss terms, while the fused loss keeps the combined embedding well-conditioned.

\subsection{Architecture specification}\label{app:architecture_spec}

Table~\ref{tab:architecture} lists the full specification of the final model. The light curve encoder is a two-layer Transformer with a model dimension of 64, two attention heads, and a feed-forward dimension of 128, totalling 83\,968 parameters. Temporal information enters through 32 learnable Fourier frequency-phase pairs, and the variable-length output sequence is compressed to a fixed 64-dimensional vector by learned attention pooling with a single query. The periodogram encoder is a one-dimensional ResNet with a wide-kernel stem followed by four residual blocks that progressively widen the channel count from 16 to 64, totalling 85\,120 parameters. Both branches output 64 dimensions, which are concatenated and mapped to the final 64-dimensional embedding by a single fusion layer with 8\,256 parameters. The full model contains 177\,344 trainable parameters.

\begin{table}[!ht]
\caption{Architecture specification of the final model.}
\label{tab:architecture}
\centering
\begin{tabular}{l l}
\hline\hline
\multicolumn{2}{l}{\textit{Light curve encoder (Transformer)}} \\
\hline
Input features & 2 (time, magnitude) \\
Time encoding & Learnable Fourier (32 freq.) \\
Model dimension & 64 \\
Attention heads & 2 \\
Encoder layers & 2 \\
Feed-forward dimension & 128 \\
Activation & GELU \\
Normalisation & LayerNorm (pre-norm) \\
Pooling & Learned attention (1 query) \\
Output dimension & 64 \\
\hline
\multicolumn{2}{l}{\textit{Periodogram encoder (ResNet)}} \\
\hline
Input & 1000 bins \\
Stem & Conv1d($1 \to 16$, $k{=}7$, $s{=}2$) \\
ResBlock 1 & $16 \to 32$, $s{=}2$ \\
ResBlock 2 & $32 \to 64$, $s{=}2$ \\
ResBlock 3 & $64 \to 64$, $s{=}2$ \\
ResBlock 4 & $64 \to 64$, $s{=}2$ \\
Normalisation & GroupNorm \\
Output pooling & AdaptiveMaxPool1d \\
Output dimension & 64 \\
\hline
\multicolumn{2}{l}{\textit{Projection and fusion}} \\
\hline
LC projection & Identity ($64 \to 64$) \\
PG projection & Identity ($64 \to 64$) \\
Fusion & Linear($128 \to 64$) + GELU \\
Embedding dimension & 64 \\
\hline
\multicolumn{2}{l}{\textit{Parameters}} \\
\hline
LC encoder & 83\,968 \\
PG encoder & 85\,120 \\
Fusion & 8\,256 \\
Total & 177\,344 \\
\hline
\end{tabular}
\tablefoot{Both halves of each source are processed by the same shared-weight encoders. Dropout of 0.15 is applied in both branches.}
\end{table}

\subsection{Training configuration and hardware}\label{app:implementation}

The model is trained using the AdamW optimiser \citep{Loshchilov2019} with a learning rate of $10^{-4}$ and weight decay of $10^{-4}$. We use a batch size of 256 and clip gradient norms to 1.0. Dropout of 0.15 is applied in both encoder branches. The learning rate is reduced by a factor of 0.5 after 3 epochs without improvement in the validation loss, and training is stopped early after 6 epochs without improvement. Class imbalance is mitigated through sqrt-inverse sampling weights (Sect.~\ref{sec:implementation}). The final model was trained for 20 epochs.

All experiments were conducted on a single ASUS Zephyrus G14 laptop with 32\,GB of system memory and an NVIDIA RTX 5070~Ti GPU (12\,GB). Training the final model for 20 epochs with on-the-fly periodogram computation at 200\,000 grid points takes approximately 87 minutes. The lightweight model and moderate batch size were deliberately chosen to keep training feasible on consumer hardware, demonstrating that this framework does not require high-performance computing resources.

\section{Embedding dimensionality}\label{app:dimension}

Table~\ref{tab:dimension} reports the classification performance as a function of the embedding dimensionality, evaluated through five-fold cross-validation on the training and validation set at the 14-class level. Performance increases with dimensionality up to 128 dimensions, where the LR reaches 77.4\% balanced accuracy. However, the SVM$-$LR gap provides additional insight. At 64 dimensions, the gap is the smallest (1.3 percentage points), indicating that the embedding is nearly fully linearly separable. At 128 dimensions, the gap increases to 1.7 points, and the SVM improves more than the LR, meaning the additional dimensions have begun to encode non-linear features that require a non-linear classifier to exploit. Since the fusion layer maps the 128-dimensional concatenation of the two branch outputs to the final embedding, a 128-dimensional output performs no compression and allows more complex, potentially less interpretable structure to pass through. The 64-dimensional embedding achieves the best balance between classification performance and linear separability, and we adopted it for this study. Below 64 dimensions, both accuracy and linear separability degrade rapidly, with gaps exceeding 3 percentage points.

\begin{table}[!ht]
\caption{Effect of embedding dimensionality.}
\label{tab:dimension}
\centering
\begin{tabular}{r c c c}
\hline\hline
$d_\mathrm{embed}$ & LR & SVM & SVM $-$ LR \\
\hline
128 & 77.4 & 79.1 & $+1.7$ \\
 64 & 76.2 & 77.5 & $+1.3$ \\
 32 & 72.7 & 76.1 & $+3.4$ \\
 16 & 68.7 & 73.0 & $+4.3$ \\
  8 & 60.7 & 64.1 & $+3.4$ \\
  4 & 57.2 & 61.4 & $+4.2$ \\
\hline
\end{tabular}
\tablefoot{Balanced accuracy at the 14-class level, per cent, from five-fold cross-validation on the training and validation set.}
\end{table}

\section{Preprocessing validation}\label{app:preprocessing}

\begin{figure}[!h]
  \centering
  \includegraphics[width=\columnwidth]{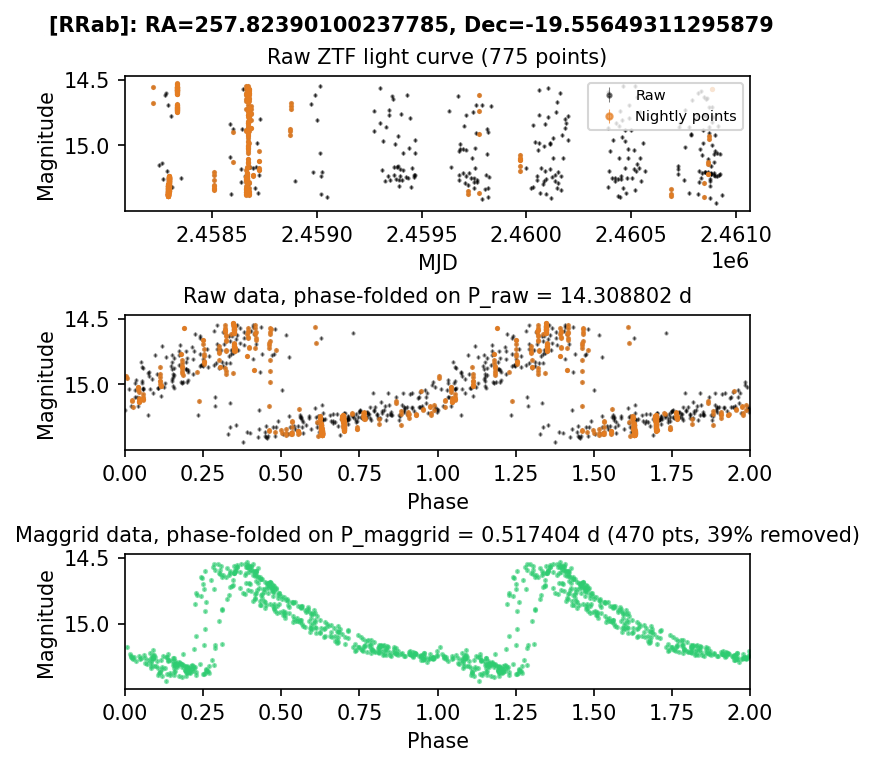}\\[6pt]
  \includegraphics[width=\columnwidth]{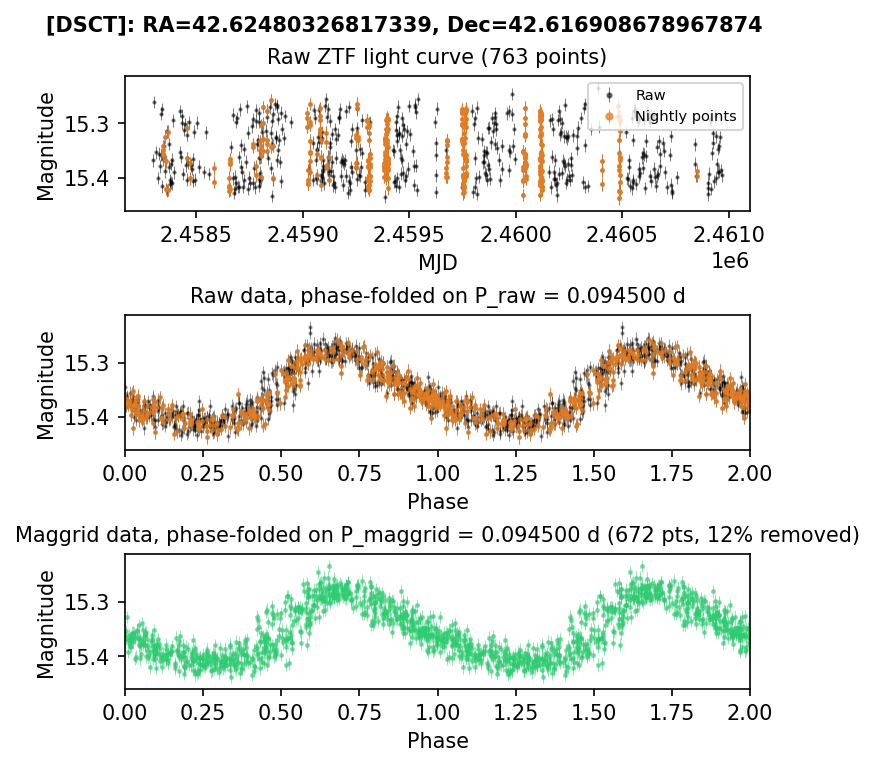}
  \caption{Effect of magnitude-grid nightly aggregation on two periodic variables. Each panel shows, from top to bottom: the raw light curve, the raw data phase-folded on the GLS period from the raw data, and the aggregated data phase-folded on the GLS period obtained after aggregation. \textit{Top:} An RR~Lyrae variable where 39\% of the points are removed. \textit{Bottom:} A $\delta$~Scuti variable where 12\% of the observations are removed.}
  \label{fig:maggrid}
\end{figure}

The magnitude-grid nightly aggregation strategy and the grid width of 0.01~mag adopted in Sect.~\ref{sec:preprocessing} were validated using a sample of 25\,088 pulsating variables from \citet{Szklenar2026} with known OGLE reference periods. Table~\ref{tab:period_recovery} compares the period recovery accuracy of the GLS periodogram on raw ZTF light curves versus magnitude-grid aggregated data at several grid widths. Aggregation consistently improves exact period recovery over the raw data (73.7\%), with the best results at grid widths of 0.01--0.05~mag (77.5--78.0\%). We adopted the 0.01~mag grid, which achieves 77.5\% exact recovery while retaining more observations per source than the 0.05~mag grid (median 679 vs 577 points). This trade-off favours the split-half augmentation described in Sect.~\ref{sec:splithalf}, where more observations per source provide richer and more diverse temporal subsets.

\begin{table}[!ht]
  \caption{Period recovery accuracy on pulsating variables.}
  \label{tab:period_recovery}
  \centering
  \begin{tabular}{l r r r}
    \hline\hline
    Preprocessing  & Exact & Harmonic & Any \\
    \hline
    Raw ZTF        & 73.7\% & 3.0\% & 76.7\% \\
    Maggrid 0.005  & 76.6\% & 2.4\% & 79.1\% \\
    Maggrid 0.01   & 77.5\% & 2.2\% & 79.7\% \\
    Maggrid 0.05   & 78.0\% & 1.9\% & 79.9\% \\
    Maggrid 0.10   & 77.6\% & 1.8\% & 79.4\% \\
    \hline
  \end{tabular}
  \tablefoot{Based on 25\,088 pulsating variables from \citet{Szklenar2026} with OGLE reference periods. Exact: GLS peak period matches the OGLE period within 0.1\%. Harmonic: GLS peak matches a harmonic of the OGLE period. Any: exact or harmonic match.}
\end{table}

\FloatBarrier
\section{Dense periodogram grid resolution}\label{app:ndense}

During the development of the model the dense periodogram grid was fixed at 300\,000 points (Sect.~\ref{sec:periodogram_computation}), an oversampling chosen so that frequency resolution would never constrain any architectural or training decision. Once those decisions were finalised by the label-free selection of Sect.~\ref{sec:evaluation}, we asked how coarse the grid could be made without sacrificing the information the representation relies on. We retrained the model at a range of grid resolutions and evaluated each by five-fold cross-validation on the training and validation set. Table~\ref{tab:ndense} reports the balanced accuracy at the 14- and 12-class levels together with the training time relative to the coarsest grid, since the periodogram is recomputed on the fly during training and a finer grid therefore makes every epoch slower.

\begin{table}[!ht]
\caption{Effect of the dense periodogram grid resolution.}
\label{tab:ndense}
\centering
\begin{tabular}{r c c c}
\hline\hline
$N_\mathrm{grid}$ & LR$_{14}$ & LR$_{12}$ & $t_\mathrm{rel}$ \\
\hline
1\,000    & 65.8 & 72.2 & 1.00 \\
10\,000   & 67.6 & 75.1 & 1.06 \\
25\,000   & 71.0 & 78.1 & 1.20 \\
50\,000   & 72.8 & 79.9 & 1.53 \\
100\,000  & 75.7 & 82.8 & 3.11 \\
200\,000  & 76.4 & 83.4 & 6.42 \\
300\,000  & 76.4 & 83.2 & 10.45 \\
\hline
\end{tabular}
\tablefoot{Balanced accuracy (per cent) from 5-fold CV on the training and validation set, at the 14- and 12-class levels, for the LR classifier. $t_\mathrm{rel}$ is the total training time for 20 epochs with on-the-fly periodogram computation at each resolution, relative to the 1000-point grid, on the hardware listed in Appendix~\ref{app:implementation}.}
\end{table}

Performance rises steeply up to roughly $10^5$ points and then saturates. The 200\,000- and 300\,000-point grids are statistically indistinguishable, at 76.4 per cent at the 14-class level for both, so nothing is gained by going finer than 200\,000, which we adopt for the released model. Reducing the resolution to 100\,000 points costs less than one point of accuracy, at 75.7 per cent, while roughly halving the training time relative to the adopted grid, which makes it an attractive setting when compute is the binding constraint. Below $10^5$ points the loss of narrow-peak resolution becomes damaging, with the coarsest grids shedding ten points or more. These are independent training runs from the main model, so their absolute accuracies differ from those of Sect.~\ref{sec:res_classification} by a few tenths of a point through random initialisation, and it is the trend with resolution that is the robust result.

\section{Per-class modality comparison}\label{app:modality_perclass}

Table~\ref{tab:modality_perclass} reports the per-class recall for the combined, light curve-only, and periodogram-only models evaluated through five-fold cross-validation on the training and validation set with the logistic regression classifier. The rightmost column shows the gain of the combined model over the better-performing single modality for each class.

\begin{table}[!ht]
\caption{Per-class recall by input modality.}
\label{tab:modality_perclass}
\centering
\begin{tabular}{l r r r r}
\hline\hline
Class & LC & PG & Comb. & $\Delta$ \\
\hline
CEP     & 38.2 & 60.2 & 75.1 & $+14.9$ \\
DSCT    & 85.5 & 85.9 & 98.0 & $+12.1$ \\
YSO     & 35.8 & 26.0 & 47.3 & $+11.5$ \\
LPV     & 16.6 & 64.5 & 73.3 & $+8.8$ \\
SN~Ia    & 67.9 & 74.2 & 82.4 & $+8.2$ \\
Blazar  & 47.5 & 16.2 & 55.0 & $+7.5$ \\
RRab   & 81.9 & 43.8 & 87.6 & $+5.7$ \\
RRc    & 86.5 & 55.7 & 92.0 & $+5.5$ \\
AGN     & 46.0 & 30.8 & 49.1 & $+3.1$ \\
QSO     & 19.1 & 62.0 & 64.2 & $+2.2$ \\
EB/EW   & 95.7 & 80.6 & 97.6 & $+1.9$ \\
CV/Nova & 77.7 & 60.2 & 79.4 & $+1.7$ \\
RS~CVn   & 87.1 & 42.4 & 85.8 & $-1.3$ \\
EA      & 90.4 & 50.7 & 89.0 & $-1.4$ \\
\hline
\end{tabular}
\tablefoot{Balanced recall per class, per cent, from 5-fold CV on the training and validation set with the LR classifier, sorted by the gain $\Delta$ of the combined model over the better-performing single modality.}
\end{table}

The geometric variables (EA and EB/EW) achieve high recall with the light curve branch alone (90.4\% and 95.7\%), as their distinctive eclipse shapes are directly visible in the time-domain data. Adding the periodogram provides little or no improvement for these classes. The pulsating variables show the opposite pattern. Cepheids improve from 38.2\% (LC) and 60.2\% (PG) to 75.1\% (combined), and $\delta$~Scuti variables from 85.5\% (LC) and 85.9\% (PG) to 98.0\% (combined), demonstrating that the two modalities contribute complementary information for period-dependent classes. LPVs benefit substantially from combining both modalities, improving from 16.6\% (LC) and 64.5\% (PG) to 73.3\% (combined). The light curve branch alone struggles with LPVs, likely because the long pulsation periods are difficult to recognise from unfolded, irregularly sampled data without frequency-domain information. The stochastic classes (AGN, Blazar, QSO) and YSOs show moderate gains from combining modalities. Type~Ia supernovae improve from 67.9\% (LC) and 74.2\% (PG) to 82.4\% (combined), indicating that both the temporal profile and the frequency-domain signature of the transient contribute to their identification.

\section{Extended correlation analysis}\label{app:correlations}

Table~\ref{tab:correlations_extended} extends the correlation analysis of Sect.~\ref{sec:res_properties} with ten additional features computed using \texttt{feets} \citep{Cabral2018}. These features are grouped by the light curve property they measure, and features within each group consistently map to the same embedding dimension, confirming that the representation organises information along physically interpretable axes.

\begin{table}[!ht]
\caption{Extended physical property correlations.}
\label{tab:correlations_extended}
\centering
\begin{tabular}{l c c}
\hline\hline
Property & Dim. & $|\rho|$ \\
\hline
Median amplitude       & 55 & 0.852 \\
Std                    & 55 & 0.836 \\
Gskew                  & 29 & 0.769 \\
Median abs. dev.       &  8 & 0.667 \\
Reduced $\chi^2$       & 40 & 0.644 \\
Roms                   & 40 & 0.639 \\
Weighted beyond 1$\sigma$ & 41 & 0.625 \\
Interperc. range (25)  &  8 & 0.608 \\
CAR $\tau$             & 43 & 0.503 \\
Stetson $K$            & 17 & 0.435 \\
\hline
\end{tabular}
\tablefoot{Same format as Table~\ref{tab:correlations}. All features computed using \texttt{feets} \citep{Cabral2018}.}
\end{table}

Variability strength is encoded primarily in embedding dimension 55. Both the standard deviation (Std, $|\rho| = 0.836$) and the median amplitude ($|\rho| = 0.852$) correlate strongly with this dimension, which is distinct from dimension 39 where the peak-to-peak amplitude resides. The embedding thus separately encodes statistical spread and peak-to-peak range. The generalised skewness Gskew ($|\rho| = 0.769$) maps to dimension 29, independently validating the asymmetry captured by the $M$ index on the same dimension. Deviation from constancy is captured by dimension 40, where the reduced $\chi^2$ ($|\rho| = 0.644$) and the robust median statistic Roms ($|\rho| = 0.639$) both reside alongside the harmonic amplitude features (not listed individually). The median absolute deviation ($|\rho| = 0.667$) and the interpercentile range ($|\rho| = 0.608$) map to dimension 8, confirming the robust spread information captured by $Q_{31}$. The fraction of observations beyond one weighted standard deviation ($|\rho| = 0.625$) maps to its own dimension (41), capturing the prevalence of outlying observations. The characteristic timescale from a continuous autoregressive model (CAR $\tau$, $|\rho| = 0.503$) maps to dimension 43, independently validating the autocorrelation length on the same dimension. Stetson~$K$ ($|\rho| = 0.435$), a kurtosis-like index widely used in variable star studies, maps to dimension 17.

Notably, the harmonic relative phases computed by \texttt{feets} show correlations below $|\rho| = 0.05$ across all embedding dimensions, indicating that the model correctly ignores the arbitrary observation epoch. Similarly, observation timing properties such as the mean and standard deviation of the timestamps show only weak correlations ($|\rho| < 0.17$), confirming that the embedding captures variability properties rather than observational cadence.

\begin{figure}[!h]
  \centering
  \includegraphics[width=\columnwidth]{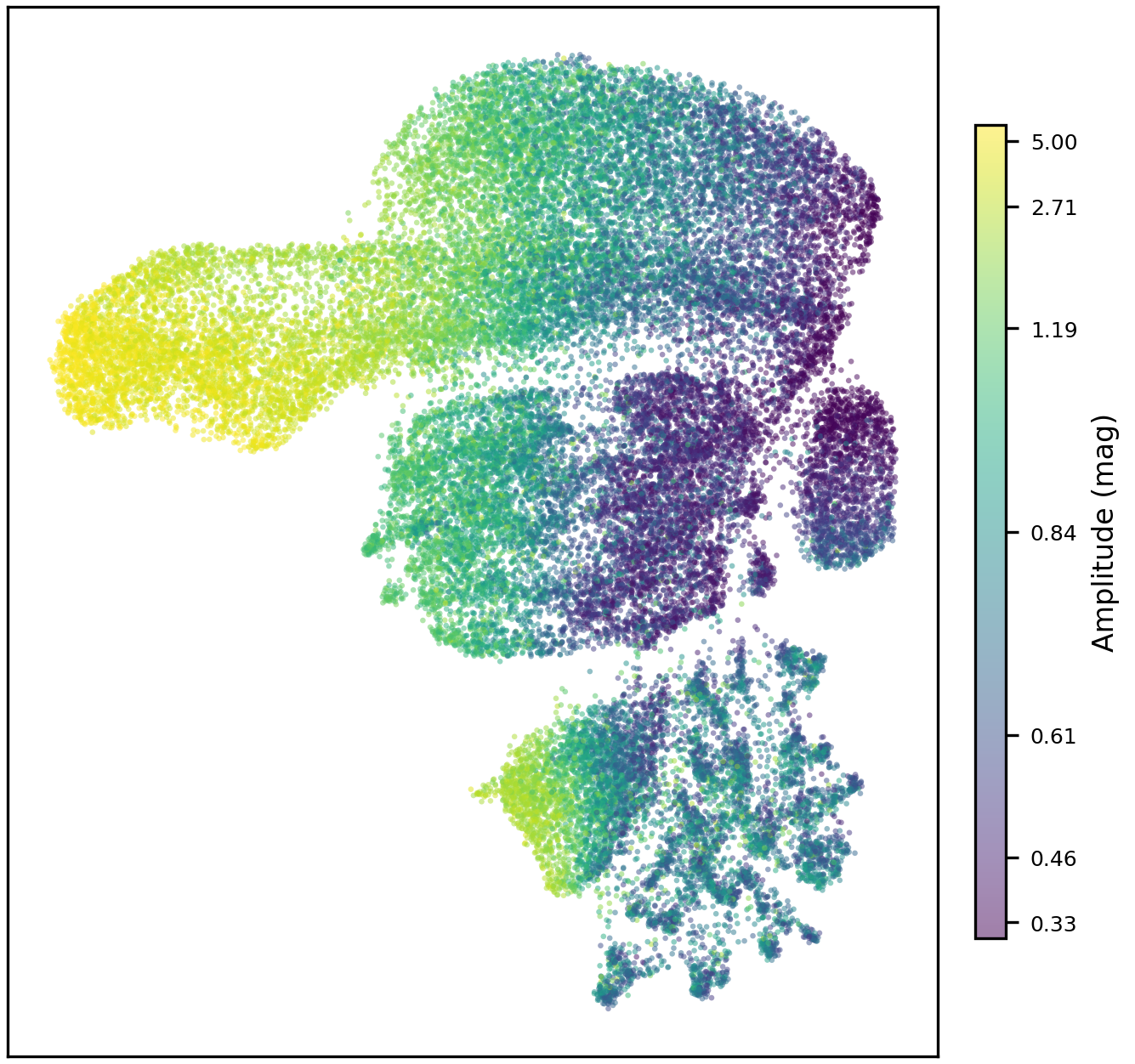}
  \caption{UMAP projection coloured by variability amplitude (histogram-equalised).}
  \label{fig:umap_amplitude}
\end{figure}

\begin{figure}[!h]
  \centering
  \includegraphics[width=\columnwidth]{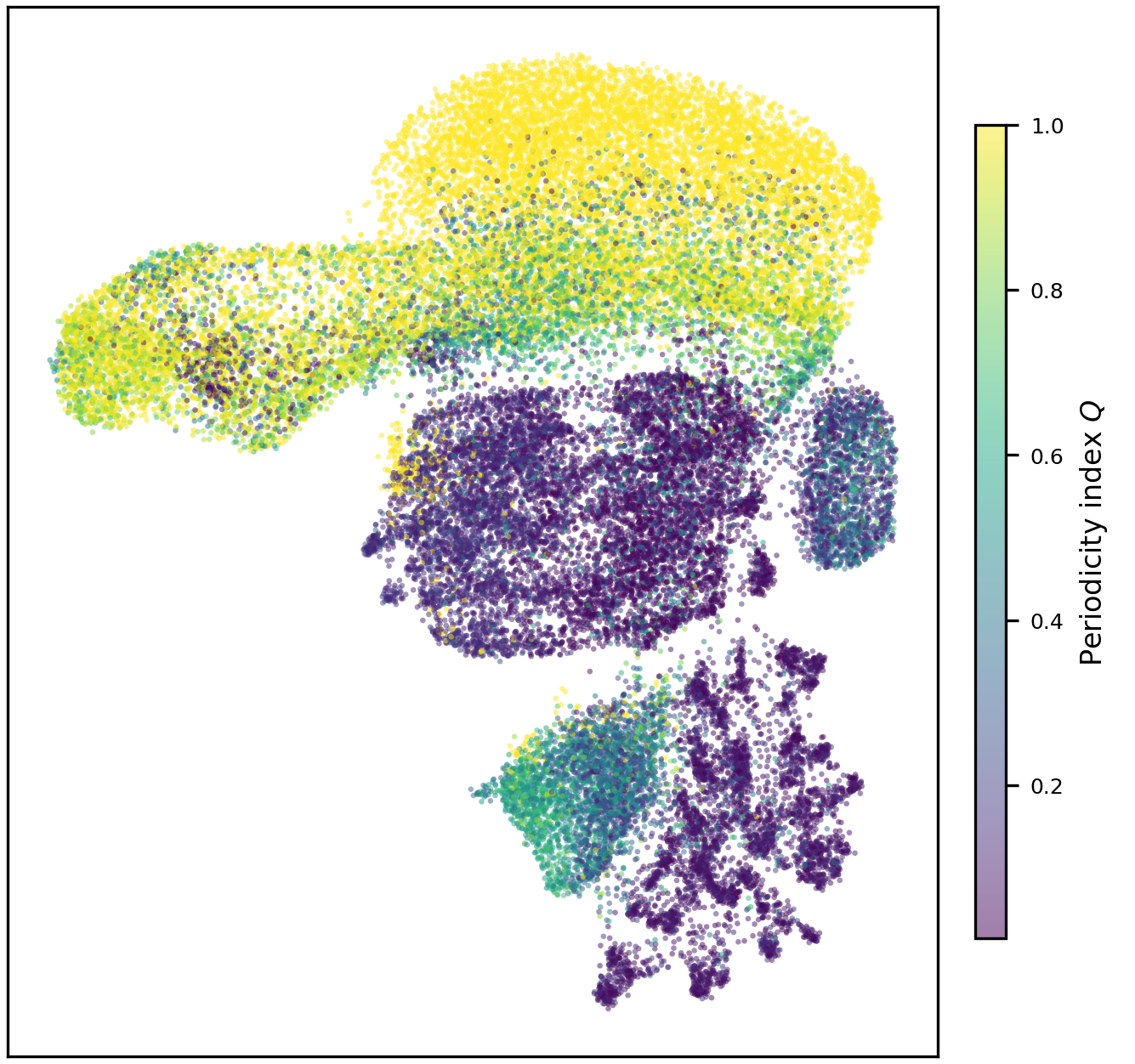}
  \caption{UMAP projection coloured by periodicity index $Q$ \citep{Cody2014}.}
  \label{fig:umap_q}
\end{figure}

\begin{figure}[!h]
  \centering
  \includegraphics[width=\columnwidth]{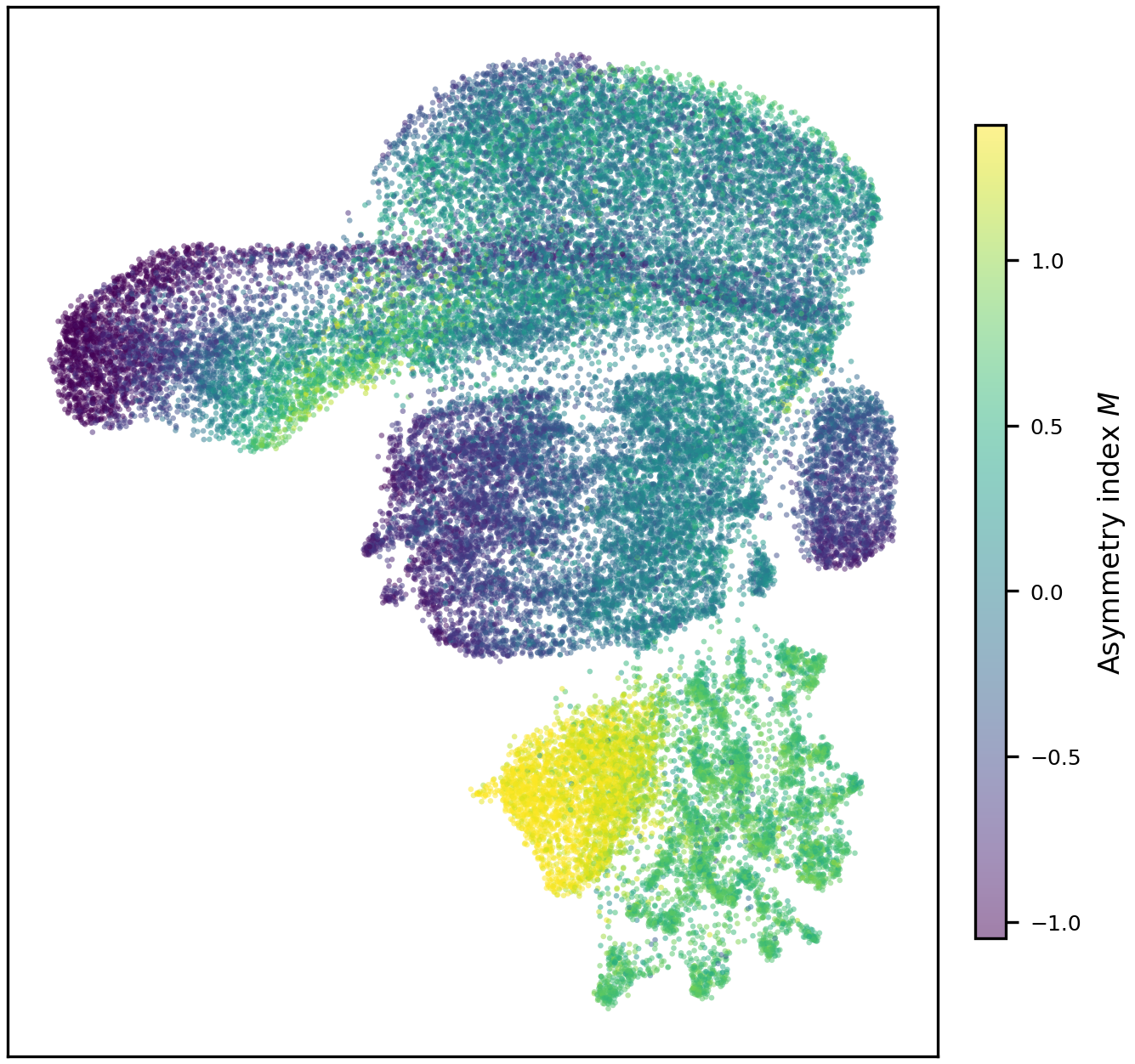}
  \caption{UMAP projection coloured by asymmetry index $M$ \citep{Cody2014}.}
  \label{fig:umap_m}
\end{figure}

\begin{figure}[!h]
  \centering
  \includegraphics[width=\columnwidth]{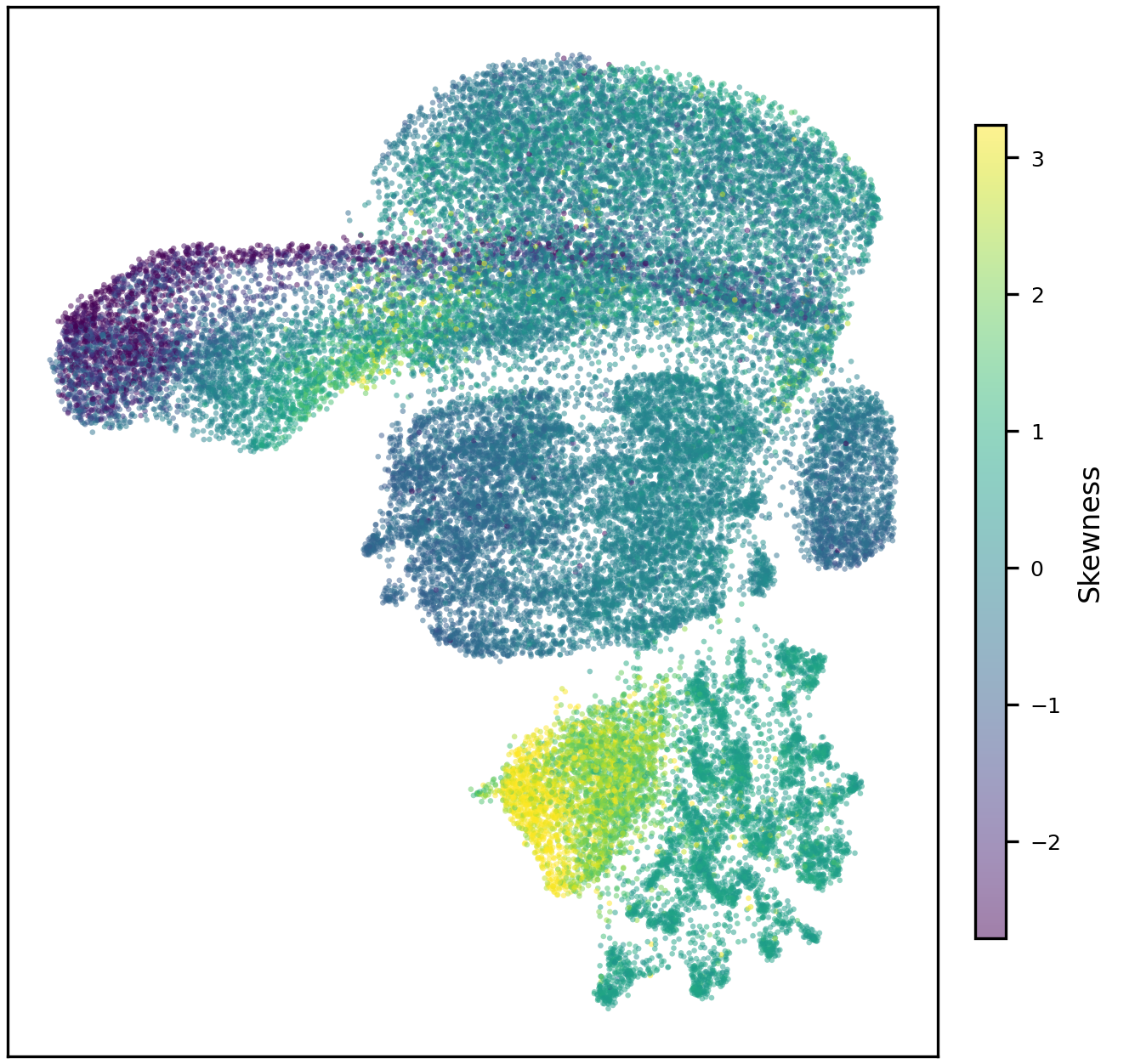}
  \caption{UMAP projection coloured by skewness of the magnitude distribution.}
  \label{fig:umap_skewness}
\end{figure}

\FloatBarrier
\section{Discussion figures}\label{app:disc_figures}

\begin{figure}[!h]
  \centering
  \includegraphics[width=\columnwidth]{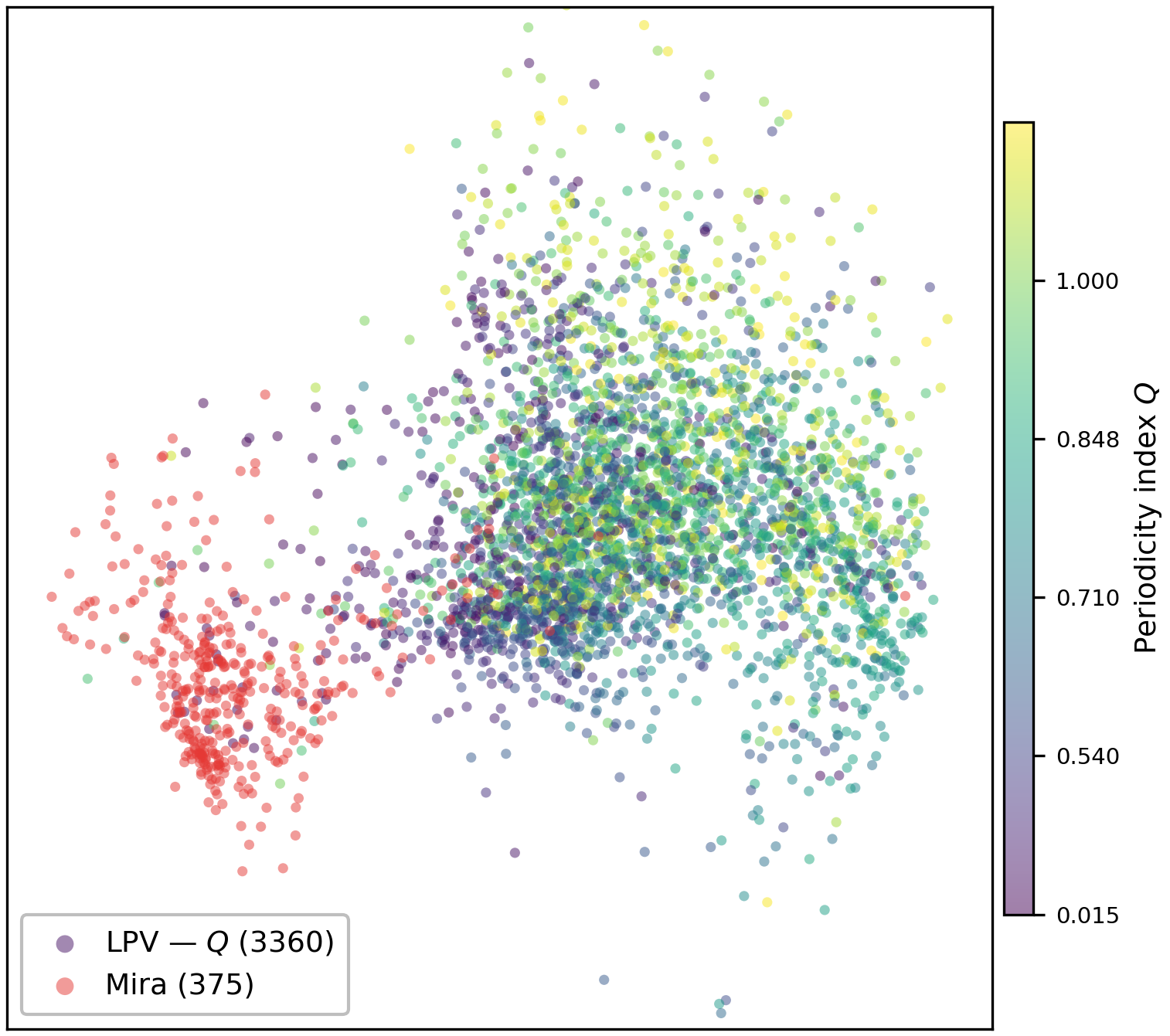}
  \caption{UMAP projection of the LPV population coloured by SIMBAD subtype. The separated subcluster on the left consists predominantly of Mira variables.}
  \label{fig:umap_lpv}
\end{figure}

\begin{figure}[!h]
  \centering
  \includegraphics[width=\columnwidth]{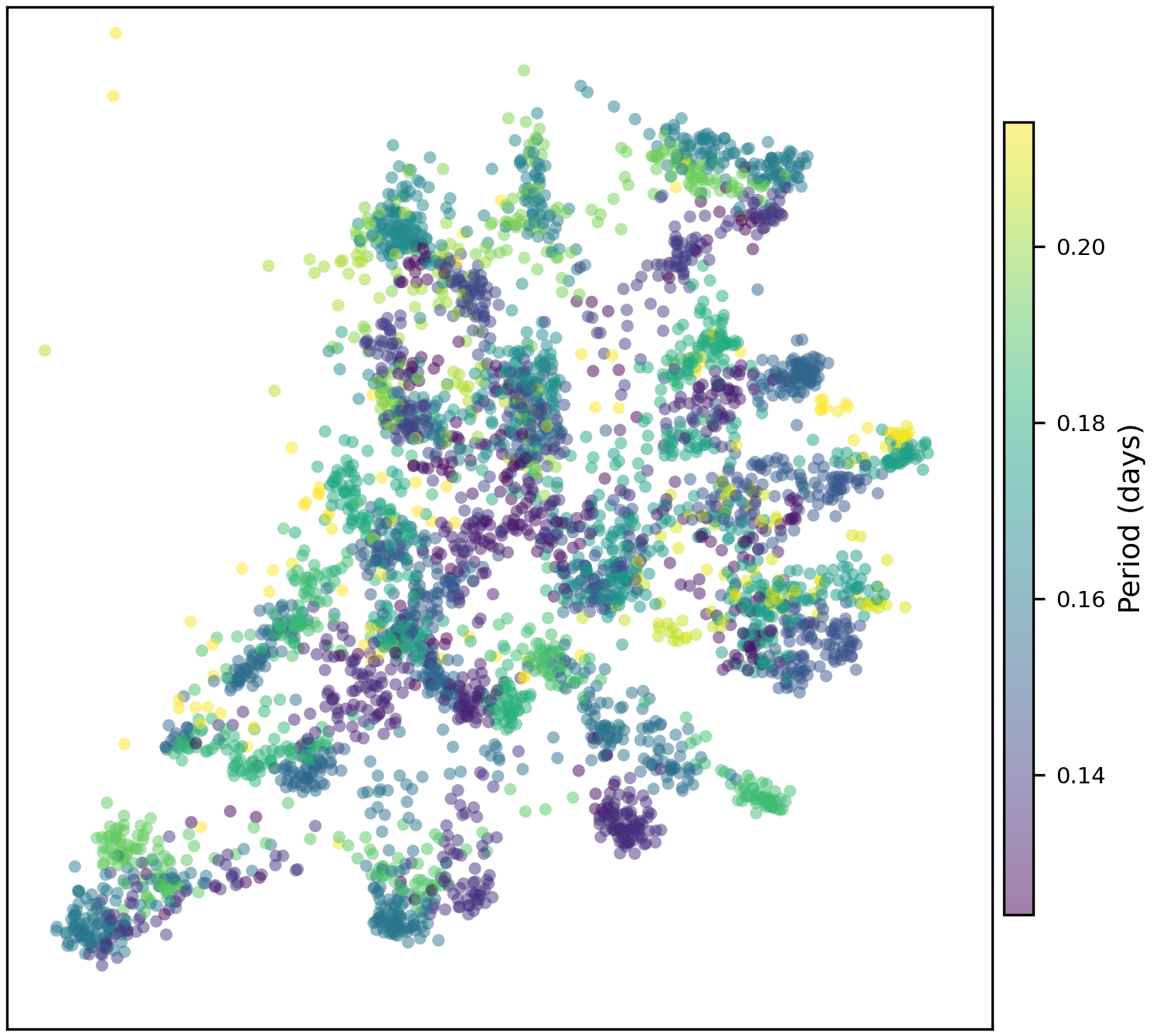}
  \caption{UMAP projection of the EB/EW population coloured by GLS period. The subclusters correspond to narrow ranges of orbital periods.}
  \label{fig:umap_ebew}
\end{figure}

\begin{figure}[!h]
  \centering
  \includegraphics[width=\columnwidth]{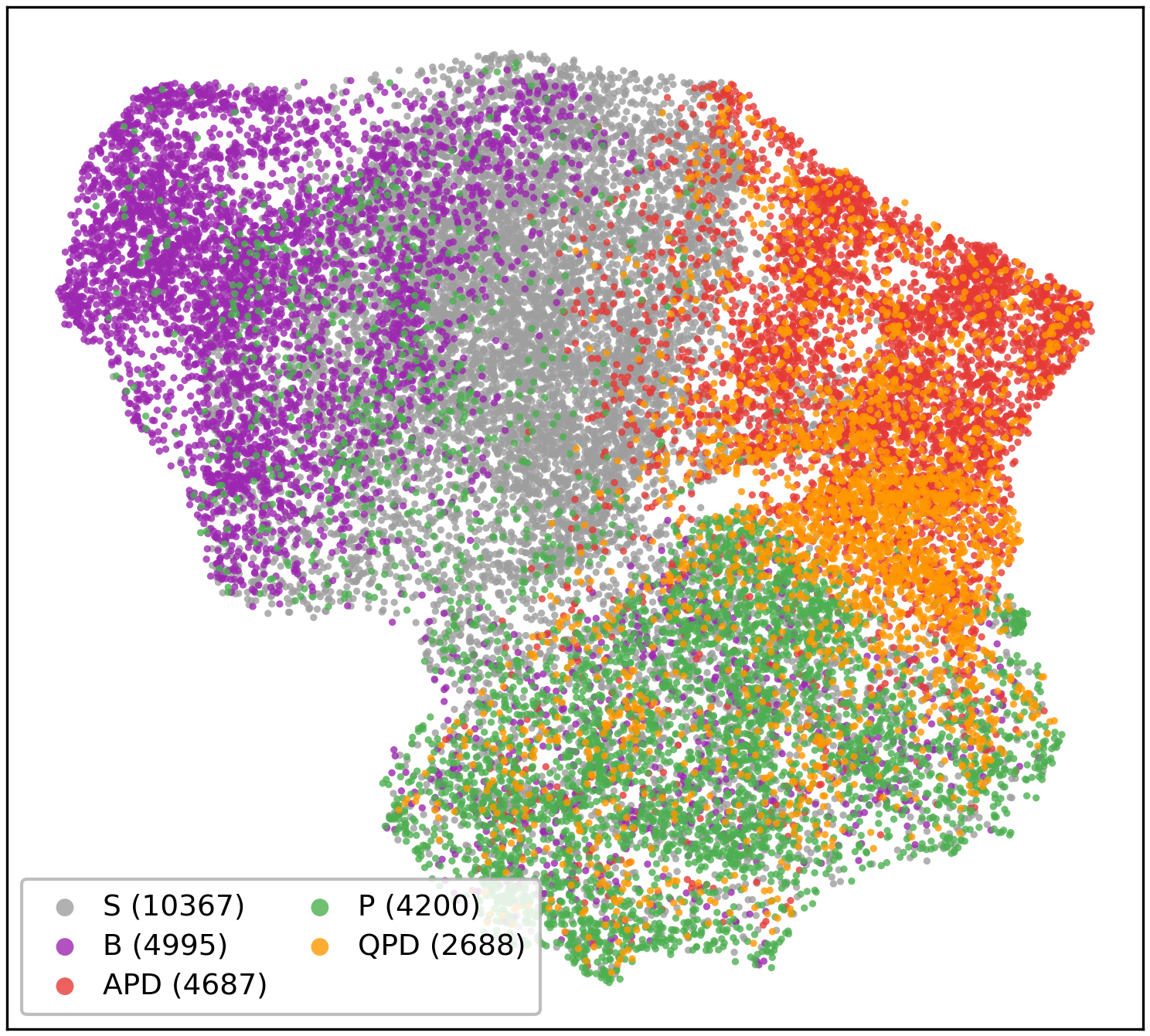}
  \caption{UMAP projection of approximately 27\,000 YSO candidates from the NGYSO catalogue \citep{Marton2026_ngyso}, coloured by morphological class following \citet{Cody2014}.}
  \label{fig:umap_yso}
\end{figure}

Figures~\ref{fig:umap_lpv}--\ref{fig:umap_yso} supplement the discussion in Sect.~\ref{sec:discussion}.

\FloatBarrier
\section{Multi-band per-class results}\label{app:multiband}

Table~\ref{tab:multiband_perclass} reports the per-class recall for all single-band and concatenated multi-band configurations evaluated through 5-fold CV with an LR classifier on the 17\,442 sources common to all three ZTF bands. The three single-band models were trained independently on their respective bands and differ only in their input light curves.

\begin{table}[!ht]
\caption{Per-class recall by single-band and concatenated multi-band embeddings.}
\label{tab:multiband_perclass}
\centering
\begin{tabular}{l r r r r r r r r}
\hline\hline
Class & $N$ & $g$ & $r$ & $i$ & $g$+$r$ & $g$+$i$ & $r$+$i$ & $g$+$r$+$i$ \\
\hline
AGN     & 1528 & 37.3 & 46.2 & 52.7 & 62.7 & 63.5 & 57.5 & 68.8 \\
Blazar  &  201 & 56.2 & 55.2 & 52.2 & 56.7 & 60.7 & 58.7 & 61.7 \\
CEP     &  273 & 76.6 & 68.1 & 69.2 & 80.6 & 84.2 & 78.0 & 83.2 \\
CV/Nova & 1317 & 79.1 & 80.0 & 74.5 & 82.4 & 80.9 & 80.6 & 83.5 \\
DSCT    &  825 & 98.3 & 96.8 & 93.2 & 99.2 & 99.4 & 98.5 & 99.4 \\
EA      &  897 & 90.5 & 89.3 & 86.5 & 92.0 & 91.0 & 90.1 & 91.1 \\
EB/EW   & 1259 & 98.6 & 97.5 & 93.4 & 99.4 & 99.0 & 98.6 & 99.8 \\
LPV     &  951 & 48.1 & 72.2 & 83.2 & 78.4 & 87.9 & 89.0 & 90.7 \\
QSO     & 4157 & 68.4 & 62.6 & 50.7 & 76.4 & 75.2 & 66.3 & 79.1 \\
RRab   & 2433 & 94.6 & 87.3 & 82.4 & 95.2 & 95.5 & 90.5 & 96.3 \\
RRc    & 2184 & 94.2 & 90.3 & 83.9 & 95.2 & 95.4 & 93.0 & 96.1 \\
RS~CVn   &  135 & 87.4 & 83.0 & 77.8 & 84.4 & 88.1 & 84.4 & 85.9 \\
SN~Ia    &  941 & 87.7 & 86.5 & 73.4 & 92.1 & 93.5 & 90.9 & 93.9 \\
YSO     &  341 & 49.0 & 39.9 & 59.5 & 57.5 & 73.0 & 73.6 & 76.2 \\
\hline
BAcc    &      & 76.1 & 75.4 & 73.8 & 82.3 & 84.8 & 82.1 & 86.1 \\
\hline
\end{tabular}
\tablefoot{Recall per class (per cent) and balanced accuracy (BAcc) from 5-fold CV with an LR classifier on the 17\,442 common sources. $N$ is the number of sources per class in the common set.}
\end{table}

\end{appendix}
\end{document}